\documentclass[12pt,aps,preprint,nofootinbib]{revtex4}

\usepackage{graphicx}
\usepackage{cancel}
\usepackage{amssymb}
\usepackage{textcomp}
\usepackage{amsmath}
\usepackage{bm}
\usepackage{times}
\usepackage{epsfig}
\usepackage{epstopdf}
\usepackage{color}
\usepackage{multirow}
\usepackage[colorlinks=true, pdfstartview=FitV, linkcolor=blue, citecolor=blue, urlcolor=blue]{hyperref}

\def\lsim{\mathrel{\raise.3ex\hbox{$<$\kern-.75em\lower1ex\hbox{$\sim$}}}}
\def\gsim{\mathrel{\raise.3ex\hbox{$>$\kern-.75em\lower1ex\hbox{$\sim$}}}}

\def\beq{\begin{equation}}
\def\eeq{\end{equation}}
\def\be{\begin{equation}}
\def\ee{\end{equation}}
\def\bea{\begin{eqnarray}}
\def\eea{\end{eqnarray}}

\newcommand{\minigraph}[5][0.25in]{\begin{minipage}{#2}\begin{center}\includegraphics[width=#2]{#5}\\\vspace{#3}\hspace{#1}{\footnotesize #4}\end{center}\end{minipage}}

\begin{document}

%\title{Revisiting Simplified Dark Matter Models in light of Gamma Ray Excess}
\title{Simplified Dark Matter Models Confront the Gamma Ray Excess}

%\vskip -1cm

\author{Csaba Bal\'azs}

\author{Tong Li}

\affiliation{
ARC Centre of Excellence for Particle Physics at the Tera-scale, School of Physics, Monash University, Melbourne, Victoria 3800, Australia
}

\begin{abstract}
Inspired by the excess of gamma rays from the Galactic Center, we confront a number of simplified dark matter models with experimental data.  Assuming a single dark matter particle coupled to standard matter via a spin-0 mediator, we compare model evidences for Majorana fermion, real scalar and real vector dark matter candidates.  We consider dark matter annihilation into various fermionic final states contributing to the observed differential gamma ray flux.  Our likelihood function also includes the dark matter relic density, its elastic scattering cross section with nuclei, and collider limits.  Using Bayesian inference we confine the mass and couplings strengths of the dark matter and mediator particle.  Our results show that, if the gamma ray excess is due to dark matter the above parameters are well constrained by the observations.  We find that the Majorana fermion dark matter model is supported the most by the data.
\end{abstract}

\maketitle

\section{Introduction}

% simplified approach
Dark matter (DM) is a mysterious component of the Universe: it constitutes nearly $25\%$ of its energy density~\cite{Ade:2013zuv} but its microscopic properties are largely unknown.  To describe these microscopic properties it is plausible to start with minimal and general theoretical assumptions.  Effective field theory (EFT) is such an approach to describe the interactions between dark and standard matter~\cite{EFT}.  In this framework all degrees of freedom, other than the single dark matter candidate, are typically assumed to be either heavy enough to be integrated out or coupling to the observable sector with negligible strength.  In some cases, however, the EFT approach to dark matter is unsatisfactory.  At a collider, or when high momentum transfer is possible, the EFT is quickly pushed to its limit of applicability~\cite{Busoni:2013lha, Goodman:2011jq, Buchmueller:2013dya}.  Avoiding such drawbacks of the EFT, a more flexible theory alternative is extending effective operators by introducing a mediator field between dark and standard matter~\cite{Berlin:2014tja,Curtin:2014afa}.

% gamma ray excess
While theoretical models of dark matter become more and more sophisticated a substantial experimental effort is dedicated to the detection dark matter particles.  One of the most promising avenues to this is indirect detection: the observation of annihilation or decay products of dark matter. Amongst the indirect signals of dark matter, the flux of gamma rays gains particular attentions~\cite{gammaray}. Over the last five years an increasingly significant deviation from a background expectation has been isolated in the measurements of the Fermi-LAT satellite~\cite{Goodenough:2009gk,Hooper:2010mq,Boyarsky:2010dr,Abazajian:2012pn,Hooper:2013rwa,Gordon:2013vta,Daylan:2014rsa}.  This deviation is apparent in high energy photons, gamma rays around 2 GeV, originating from an extended region centered on the Galactic Center.  At the moment the source of the excess photons is uncertain. They could originate from dark matter annihilation, from a population of millisecond pulsars or supernova remnants~\cite{Carlson:2014cwa}, or from cosmic rays injected in a sequence of burst-like or continuous events at the galactic center~\cite{Petrovic:2014uda}.
Based on their the spectra and luminosity function it is, however, challenging to explain the excess with millisecond pulsars~\cite{Cholis:2014lta,Cholis:2014noa}.

Recently, Daylan et al. re-analyzed data from the Large Area Telescope (LAT) on board the Fermi Gamma-Ray Space Telescope (Fermi)~\cite{Atwood:2009ez} and found excess flux of gamma rays from the direction of the Galactic Center.  They concluded that the $1\sim 3$ GeV gamma ray signal is statistically significant and appears to originate from dark matter particles annihilating rather than standard astrophysical sources~\cite{Daylan:2014rsa}.  This gamma ray excess drew the attention of a number of particle model builders and phenomenologists~\cite{Berlin:2014tja,Carlson:2014cwa,Petrovic:2014uda,AMartin,excesspapers}.

The shape of the peak in the energy spectrum is broadly consistent with gamma rays emitted from the self-annihilation of dark matter particles~\cite{Daylan:2014rsa,Gomez-Vargas:2013bea,Abazajian:2014fta,Ipek:2014gua,Ko:2014gha,Cline:2014dwa,Ko:2014loa}.  The intensity of the observed signal is consistent with an annihilation cross section that is required for a dark matter particle to freeze out with the observed abundance~\cite{Kong:2014haa,Boehm:2014bia,Ghosh:2014pwa,AMartin,Wang:2014elb,Fields:2014pia}.  Finally, the diffuse spherical nature and morphology of the excess gamma rays is consistent with a Navarro-Frenk-White-like spatial distribution of dark matter within our Galaxy~\cite{Calore:2014xka}.  For these reasons the Fermi-LAT gamma ray excess is considered by some as a smoking gun for the first indirect detection of dark matter particles.

Unfortunately, the conclusion that we discovered dark matter cannot be drawn just yet.  We not only have to try to exclude the possibility of a standard astrophysical explanation, but we also need a full concordance between the various pieces of data that point to a dark matter particle with a given mass, spin, and interaction strength to the standard sector.  To this end, we have to determine the properties of dark matter that the galactic center gamma ray excess implies and then we have to observe dark matter particles with the same properties by other means, such as direct detection or collider production.  The main aim of this paper is the first step: extraction of the microscopic properties of dark matter particles from the gamma ray excess and other existing limits on dark matter.

% % %

% this work
Previous studies of this gamma ray excess either considered specific dark matter candidates coupling only to a certain standard fermion (mostly $b\bar{b}$) and/or fixed the dark matter particle mass (typically to $\sim 35$ GeV).  By contrast, we use a model independent approach and allow the dark matter mass and couplings to vary.  We consider a single dark matter particle with a mediator that couples to various standard fermions.  Our dark matter particle thus annihilates to several final states which all contribute to the differential gamma ray flux.  We show that, under the assumption that the gamma ray excess originates from such a dark matter candidate, the data constrain the model parameters fairly well.  Our likelhood function also contains other observables such as the relic density of dark matter, the dark matter-nucleus elastic scattering cross section, and collider limits.  Using Bayesian inference, we confine the mass and couplings strengths of the dark matter and mediator particle~\cite{Balazs:2014rsa}.

% outline
The rest of the paper is organized as follows. In Sec.~\ref{sec:Models}, we summarize the simplified dark matter models we consider. % ???
In Sec.~\ref{sec:Constraints}, we discuss dark matter observables: dark matter abundance, gamma rays from the Galactic Center, direction detection and collider detection. Our numerical results are given in Sec.~\ref{sec:Results}.  We summarize our conclusions in Sec.~\ref{sec:Concl}.
We collect the formulae of Bayesian inference and likelihood functions in the Appendix.

%%%%%%%%%%%%%%%%%%%%%%%%%%%%%%%%%%%%%%%%%%%%%%%%%%%%%%%%%%%%%
\section{Theoretical Hypotheses}
\label{sec:Models}
%%%%%%%%%%%%%%%%%%%%%%%%%%%%%%%%%%%%%%%%%%%%%%%%%%%%%%%%%%%%%%%%%

In this section we describe the theoretical hypotheses which we test using the gamma ray and other data.  As stated in the introduction, we use a simplified model as a more realistic description of the interactions between dark and standard matter.  Inspired by the Higgs portal mechanism~\cite{Higgsportal,Queiroz:2014yna}, we restrict the mediator to be a spin-0 field, that is a real scalar particle $S$, in our study.  We consider three different hypotheses for the identity of the dark matter particle: real scalar ($\phi$), Majorana fermion ($\chi$), and real vector ($X$).  These dark matter candidates couple to the mediator with interactions shown in Table~\ref{tab:interaction}~\cite{Berlin:2014tja}.

\def\arraystretch{1.5}
\begin{table}[h]
\begin{tabular}{|c|c|c|c|}
\hline
Hypothesis & real scalar DM
			            & Majorana fermion DM
						& real vector DM \\
\hline
DM-mediator interaction & $\displaystyle {\cal L}_\phi \supset {\mu_\phi m_\phi\over 2}\phi^2 S$
                        & $\displaystyle {\cal L}_\chi \supset {i\lambda_\chi\over 2}\bar{\chi}\gamma_5\chi S$
						& $\displaystyle {\cal L}_X \supset {\mu_X m_X\over 2}X^\mu X_\mu S$ \\
\hline
\end{tabular}
\caption{Dark matter to mediator couplings in the three different dark matter hypothesis we consider.}
\label{tab:interaction}
\end{table}

In Table~\ref{tab:interaction} we require the interaction strengths of $\phi$ and $X$ to be proportional to the dark matter mass so that one obtains dimensionless couplings $\mu_\phi$ and $\mu_X$.  The interaction between the mediator and standard fermions $f$ is assumed to be:
\begin{eqnarray}
{\cal L}_S \supset \lambda_f \bar{f}f S.
\label{eq:intercation1}
\end{eqnarray}
In line with minimal flavor violation (MFV)~\cite{D'Ambrosio:2002ex}, we only consider the third generation standard fermions, i.e. $f=b,t,\tau$.

We assume, for simplicity, that mediator pair final states are not present in the dark matter annihilation and only consider s-channel annihilation diagrams.  According to the power counting of dark matter transfer momentum or velocity~\cite{Kumar:2013iva}, with the bi-linears in Table~\ref{tab:interaction} and Eq.~(\ref{eq:intercation1}), the annihilation cross sections of the three candidates are not velocity suppressed, i.e. $\sigma v\sim 1$.  The dark matter-nucleon elastic scattering cross section for the fermion candidate is momentum suppressed, while it is non-suppressed for the scalar and vector cases.  The resulting direct detection cross sections are all spin-independent (SI).

%%%%%%%%%%%%%%%%%%%%%%%%%%%%%%%%%%%%%%%%%%%%%%%%%%%%%%%%%%
\section{Observables}
\label{sec:Constraints}
%%%%%%%%%%%%%%%%%%%%%%%%%%%%%%%%%%%%%%%%%%%%%%%%%%%%%%%%%%%%%%%%%

%%%%%%%%%%%%%%%%%%%%%%%%%%%%%
\subsection{Dark Matter Abundance}
%%%%%%%%%%%%%%%%%%%%%%%%%%%%%%

We assume that the single dark matter candidate follows the standard thermal evolution in the early Universe and as the temperature decreases it freezes out leaving an average relic abundance.  We use micrOmegas version 3.6.9 to calculate this abundance~\cite{MO}.  Our likelihood function includes this observable, together with the experimental value determined by WMAP and Planck~\cite{Ade:2013zuv}
\begin{eqnarray}
\Omega_{DM} h^2 = 0.1199\pm 0.0027.
\end{eqnarray}
We require the relic density of the dark matter candidate to be close to the above central value, that is we use a Gaussian for this part of the likelihood function.

%%%%%%%%%%%%%%%%%%%%%%%%%%%%%
\subsection{Gamma Rays from the Galactic Center}
%%%%%%%%%%%%%%%%%%%%%%%%%%%%%%

In our scenario the gamma ray flux observed by Fermi-LAT is generated by the annihilation of self-conjugate dark matter particles.  The double-differential flux of photons as the function of energy $E_\gamma$ and observation region $\Delta \Omega$ is given by
\begin{eqnarray}
{d^2\Phi_\gamma\over dE_\gamma d\Omega}={\langle \sigma v\rangle\over 8\pi m_{DM}^2}\left(\sum_f B_f {dN_\gamma^f\over dE_\gamma}\right)J(\psi).
\label{flux}
\end{eqnarray}
Here $\langle \sigma v\rangle$ is the thermally averaged dark matter annihilation cross section, $B_f$ is the annihilation branching fraction into the $f{\bar f}$ final state, % (i.e. $B_f=\langle \sigma v\rangle_f/ \langle \sigma v\rangle$),
and $dN_\gamma^f/dE_\gamma$ is the energy spectrum of photons produced in one annihilation channel with the final state $f{\bar f}$.
The $J$ factor in Eq.~(\ref{flux}) is the function of the direction of observation $\psi$
\begin{eqnarray}
J(\psi)=\int_{los}\rho^2(r)dl,
\end{eqnarray}
with
\begin{eqnarray}
r=\sqrt{l^2+r_\odot^2-2lr_\odot \cos\psi},
\end{eqnarray}
and a generalized Navarro-Frenk-White (NFW) dark matter galactic distribution~\cite{NFW}
\begin{eqnarray}
\rho(r)=\rho_0{(r/r_s)^{-\gamma}\over (1+r/r_s)^{3-\gamma}}.
\end{eqnarray}
Here the radius of the galactic diffusion disk is $r_s=20$ kpc, the solar distance from the Galactic Center is $r_\odot=8.5$ kpc, and $\rho_0$ is set to reproduce the local dark matter density $\rho(r_\odot)=\rho_{DM}=0.3 \ {\rm GeV/cm^3}$.
Following Ref.~\cite{Daylan:2014rsa} we fix the inner slope of NFW halo profile as $\gamma=1.26$ and $\psi=5^\circ$ in order to avoid bremsstrahlung and other secondary processes~\cite{AMartin}.

\begin{figure}[t]
\begin{center}
\includegraphics[scale=1,width=8cm]{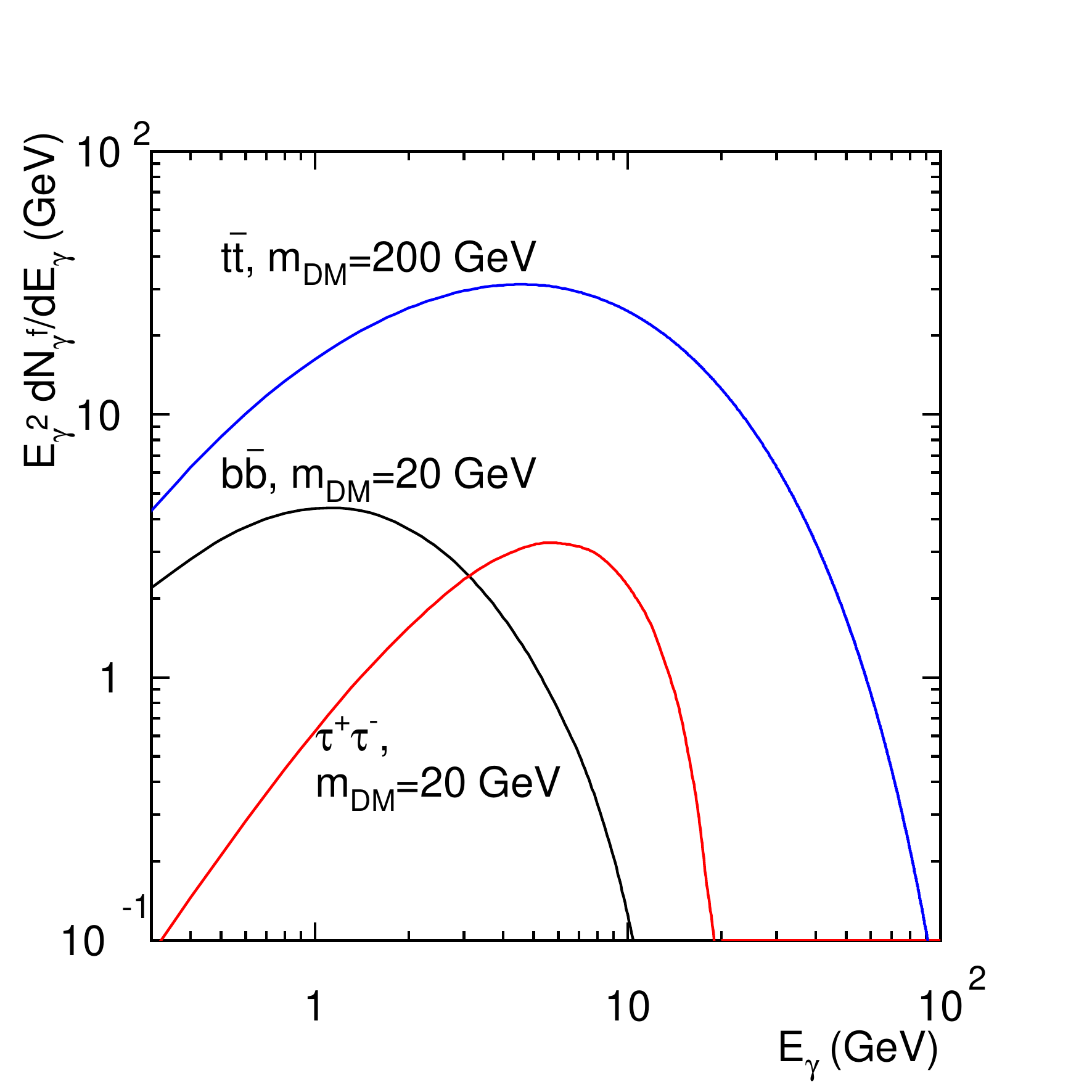}
\end{center}
\caption{The spectrum of gamma rays produced from dark matter annihilation into individual $b\bar{b},\tau^+\tau^-$ and $t\bar{t}$ final states with 20 GeV and 200 GeV dark matter mass, respectively.}
\label{yield}
\end{figure}

In Fig.~\ref{yield} we show the key component that determines the energy distribution of the gamma ray flux, that is the differential yield $dN_\gamma^f/dE_\gamma$, for the three final states we consider.  As seen from Eq.~(\ref{flux}), the differential yield is the branching fraction weighted sum of the differential yields into specific final states. Contrary to previous gamma ray studies with dark matter annihilating into only $b\bar{b}$, we sum over the contributions of the three individual Standard Model (SM) fermions ($b,t,\tau$).  As $B_f$ is model-dependent, the gamma ray data plays an important role in constraining the three different models we consider.

The gamma ray spectral data points that we input into our Gaussian likelihood function are taken from Fig.~5 in Ref.~\cite{Daylan:2014rsa}.  We use micrOmegas version 3.6.9 to evaluate the theoretical prediction for the differential gamma ray flux~\cite{MO}.

%%%%%%%%%%%%%%%%%%%%%%%%%%%
\subsection{Dark Matter Direct Detection}
%%%%%%%%%%%%%%%%%%%%%%%%%%%

As we only consider the scalar current between the mediator and the third generation quarks, the strength of the mediator-nucleon ($N$) interaction reads
\begin{eqnarray}
g_{SNN}={2\over 27}m_Nf_{TG}\sum_{f=b,t}{\lambda_f\over m_f}.
\end{eqnarray}
Above $f_{TG}=1-f_{T_u}^N-f_{T_d}^N-f_{T_s}$ and we adopt $f_{T_u}^p=f_{T_d}^n=0.02$, $f_{T_d}^p=f_{T_u}^n=0.026$, $f_{T_s}=0.043$~\cite{Berlin:2014tja,lattice,Crivellin:2013ipa}.

The elastic scattering cross section of fermionic dark matter $\chi$, as indicated in Sec. II A, is suppressed by powers of velocity~\cite{Berlin:2014tja}
\begin{eqnarray}
\sigma_{\chi N}^{SI}\simeq {\mu_{\chi N}^2 v_0^2\over 2m_\chi^2}{\mu_{\chi N}^2 \lambda_\chi^2\over \pi m_S^4}g_{SNN}^2,
\label{DDMF}
\end{eqnarray}
where $v_0$ is the mean speed of dark matter which is fixed to be $v_0=220 \ {\rm km/s}$ for simplicity, and $\mu_{\chi N}=m_\chi m_N/(m_\chi+m_N)$ is the reduced DM-nucleon mass.
The cross sections of the scalar and vector DM candidates are neither velocity nor momentum suppressed~\cite{Berlin:2014tja}:
\begin{eqnarray}
\sigma_{\phi N}^{SI}\simeq {\mu_{\phi N}^2 \mu_\phi^2\over 4\pi m_S^4}g_{SNN}^2, \ \ \ \sigma_{X N}^{SI}\simeq {\mu_{X N}^2 \mu_X^2\over 4\pi m_S^4}g_{SNN}^2.
\end{eqnarray}

For the spin dependent dark matter-proton elastic scattering cross section we use the combined upper limits of DAMIC~\cite{DAMIC}, CDMSlite~\cite{CDMSlite} and LUX~\cite{LUX} implemented in an error function shaped likelihood function.

%%%%%%%%%%%%%%%%%%%%%%%%%%%
\subsection{Collider Detection of Dark Matter}
%%%%%%%%%%%%%%%%%%%%%%%%%%%

As we only consider couplings of the mediator to the third generation quarks in our models, the most stringent collider constraint is from the direct search for dark matter associated with bottom or top quarks at the Large Hadron Collider (LHC) as studied in Ref.~\cite{Haisch:2012kf,Lin:2013sca}.  In Ref.~\cite{Lin:2013sca} searches of mono-b jet and top in pairs plus missing energy at 8 TeV LHC are analyzed to give limits on the heavy scalar mediator scale for fermionic dark matter operator.  These limits are obtained for the following effective interaction:
\begin{eqnarray}
{m_f\over M_\ast^3} \bar{\chi}\chi \bar{f}f.
\end{eqnarray}
In order to apply this limit on our models, we need to match the cut-off scale $M_\ast$ to the effective scale $M_\ast^{{\rm simp},f}$ that applies in our model, in which the latter depends on the particular final state $f{\bar f}$.  To achieve this matching we use the following relation:
\begin{eqnarray}
M_\ast^{{\rm simp},f}=\sqrt[3]{m_S^2m_f\over \lambda_\chi \lambda_f}>M_\ast^{{\rm limit},f}, \ \ \ f=b,t.
\label{lhc}
\end{eqnarray}
Here $M_\ast^{{\rm limit},f}$ is the collider limit for a particular heavy quark $f$ given in Ref.~\cite{Lin:2013sca}.

The above constraint is based on the effective field theory framework.  We have to consider the validity of effective field theory and the condition of applying the constraint in Eq.~(\ref{lhc}).  The condition for integrating out the mediator requires a mediator mass greater than the momentum transfer: $m_S > q$.
In addition, the kinematics of the mediator decaying to a pair of dark matter particles imposes $q > 2 m_\chi$, which leads to a minimal requirement on $m_S$: $m_S>2m_\chi$~\cite{Busoni:2013lha}.  This condition, however, may not be sufficient taking into account the selection cuts in Ref.~\cite{Lin:2013sca}.  We thus use $m_S>{\rm max}(2m_\chi,\cancel{E}_T^{cut})$ as the required condition in Eq.~(\ref{lhc}), where $\cancel{E}_T^{cut}$ is the cut on transverse missing energy adopted in Ref.~\cite{Lin:2013sca} with $\cancel{E}_T^{cut}=350 \ (225)$ GeV for a $b$ ($t$) quark.

Apart from the observables discussed above, there are other indirect probes which can be relevant in shaping the parameter space, for instance the measurement of radio emission and the local antiproton spectrum.
Most recently, Cholis, Hooper, and Linden critically reevaluated radio constraints on annihilating dark matter in Ref.~\cite{Cholis:2014fja}.  They found that electron energy losses are dominated by inverse Compton scattering rather than by synchrotron radiation which considerably relaxes the radio frequency constraints, while convective winds may further weaken synchrotron bounds.  Taking this into account radio constraints on annihilating dark matter are generally weaker than those derived from current gamma-ray observations.

Specifically, the left frame of Fig. 2 in Ref.~\cite{Cholis:2014fja} shows that, under reasonable assumptions on the dark matter halo profile and convection, after taking inverse Compton scattering into account the radio data do not constrain dark matter annihilation at thermal rates for $m_{DM} > 10$ GeV.  Depending on the speed of the convective wind the limits may not affect dark matter annihilating at the thermal rate down to about 1 GeV.  Since our 95 \% confidence regions are spread between $m_{DM} \simeq 10-100$ GeV for the most likely Majorana fermion case, based on Ref.~\cite{Cholis:2014fja}, we conclude that our results are not changed by radio constraints.  For the less likely real scalar and vector dark matter candidates the 95 \% confidence regions cover $m_{DM} \simeq 2.5-30$ GeV, which could be mildly affected at the low mass regions.

% positron/antiproton

Similarly, Bringmann, Vollmann, and Weniger recently updated charged cosmic-ray constraints on light dark matter in Ref.~\cite{Bringmann:2014lpa}.  The authors adapt the specific assumptions that are required by the putative dark matter signal indicated by the gamma-ray excess from the Galactic center region.  They found that cosmic-ray positron data disfavor dark matter annihilation to light leptons, or democratically to all leptons, which does not affect our conclusions.  Cosmic-ray antiprotons are in tension with dark matter annihilation to any combination of quark final states but only AMS-02 data will be able to rule out or confirm the dark matter hypothesis.

In particular, Fig. 5 in Ref.~\cite{Bringmann:2014lpa} shows that the significant uncertainty of the propagation parameters implies that antiproton data still allow for thermal dark matter annihilation into $b{\bar b}$ final states at the thermal rate for $m_{DM} \simeq 10-20$ GeV.  Fig. 4, additionally, shows that antiproton limits for the $b{\bar b}$ final state flatten out for $m_{DM} < 10$ GeV.  These two facts combined with the theoretical and experimental uncertainty in the relic abundance calculation allows us to argue that our 98 \% confidence regions may not be significantly affected by the present antiproton limits either.

%%%%%%%%%%%%%%%%%%%%%%%%%%%%%%%%%%%%%%%%%%%%%%%%%%%%%
\section{Results}
\label{sec:Results}
%%%%%%%%%%%%%%%%%%%%%%%%%%%%%%%%%%%%%%%%%%%%%%%%%%%%%

We coded the Lagrangian of the simplified dark matter models in FeynRules~\cite{FR}. Subsequent calculations, including the dark matter relic density, differential gamma ray flux, and direct detection cross section are done using micrOmegas~\cite{MO}. Nested sampling and posterior distribution calculations are performed by MultiNest~\cite{MN}. Due to their modest ranges we use log priors for all parameters. We defer details of our statistical analysis and likelihood functions to Appendix.

In the numerical calculation we fix the coupling of dark matter to the mediator as $\lambda_\chi=\mu_\phi=\mu_X=1$ and scan $m_{\chi,\phi,X},m_S$ and $\lambda_{b,t,\tau}$ in the range of $1-10^3$ GeV and $10^{-5}-10$ respectively.  We first show the posterior probability distribution taking into account of gamma ray data, relic density, and direct detection below and discuss the LHC constraint at the end.
% is only for fermionic dark matter, in order to compare our three models consistently, we do not apply such limit temporarily. We show xxxxxxxxxxxx

\begin{figure}[h]
\begin{center}
\minigraph{7cm}{-0.15in}{(a)}{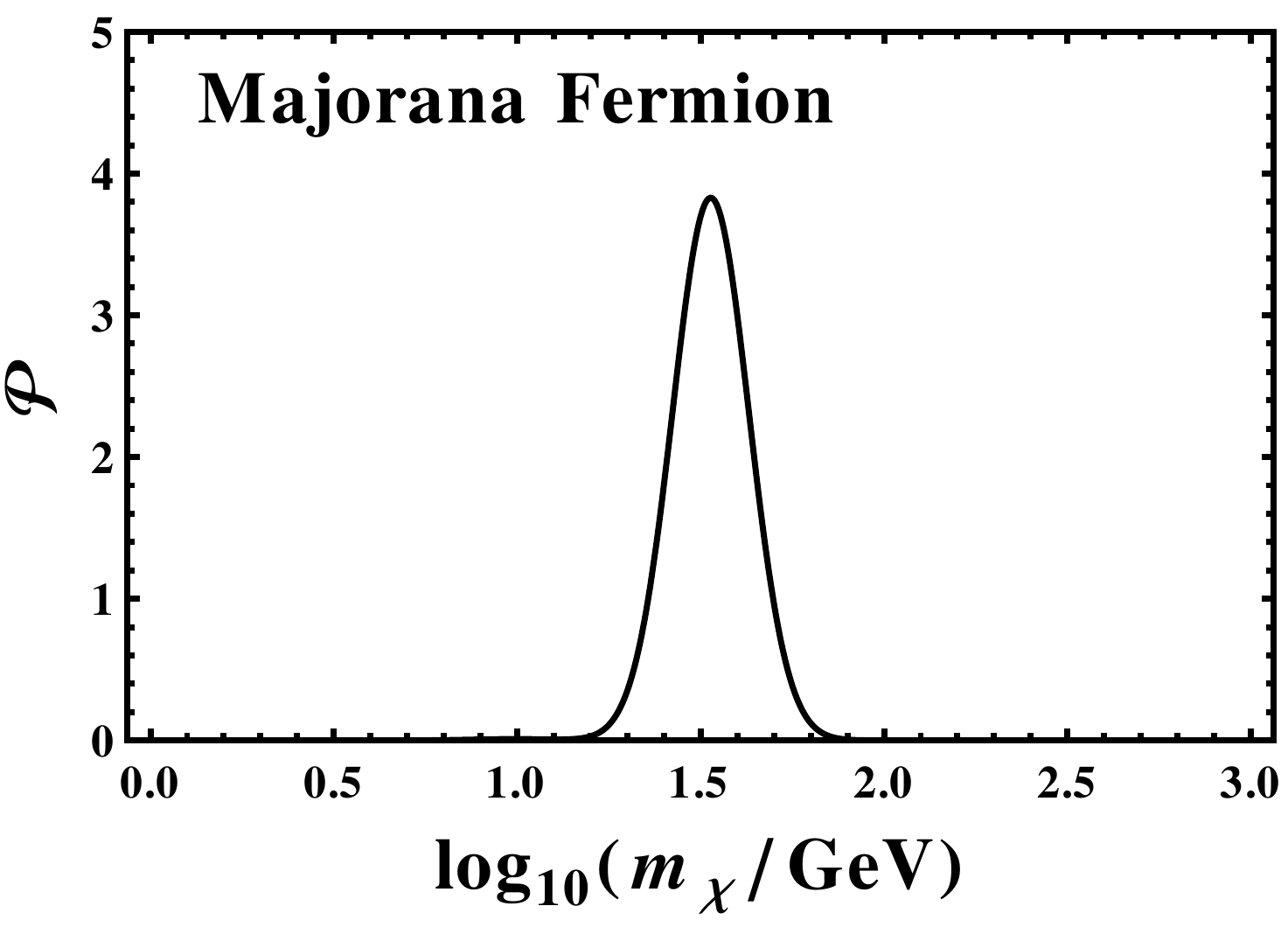}
\minigraph{7cm}{-0.15in}{(b)}{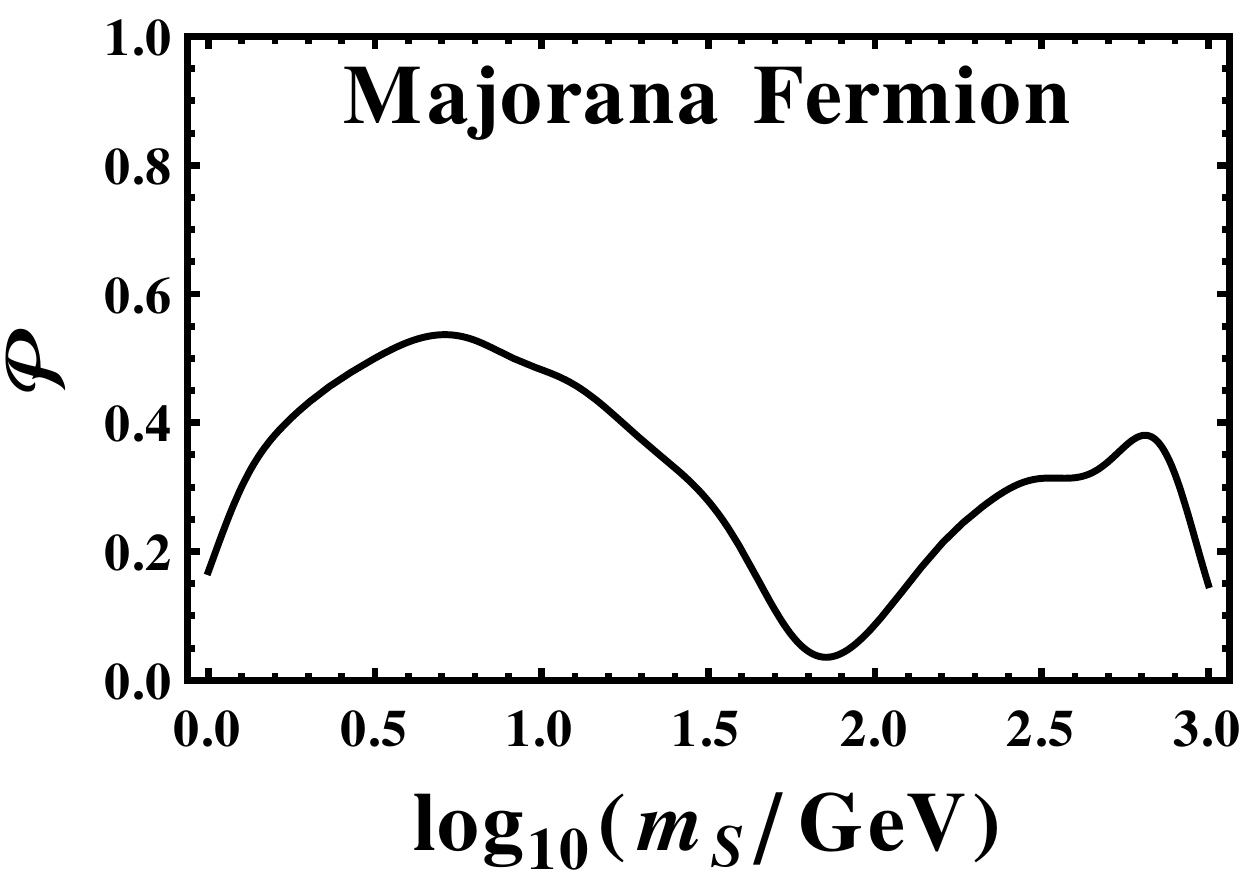}\\
\minigraph{7cm}{-0.15in}{(c)}{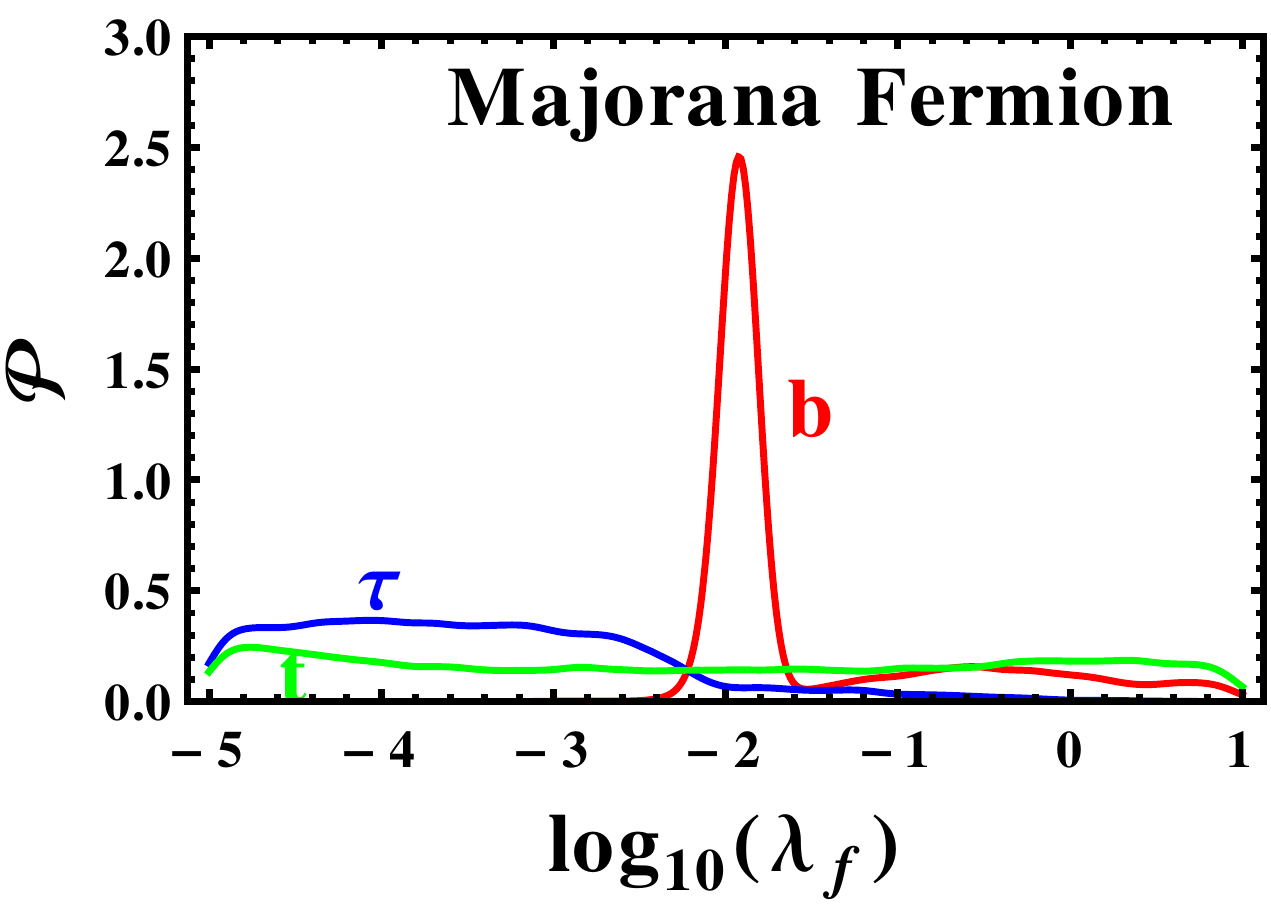}
%\minigraph{7cm}{-0.15in}{(c)}{plots/mchims-MF.pdf}
%\minigraph{7cm}{-0.15in}{(d)}{plots/SI-MF.pdf}
%\includegraphics[scale=1,width=7cm]{plots/btau-MF.pdf}
%\includegraphics[scale=1,width=7cm]{plots/bt-MF.pdf}\\
%\includegraphics[scale=1,width=7cm]{plots/mchims-MF.pdf}
%\includegraphics[scale=1,width=7cm]{plots/SI-MF.pdf}
\end{center}
\caption{Posterior probability distribution as a function of (a) $m_\chi$, (b) $m_S$ and (c) $\lambda_f$ ($f=b,t,\tau$) for Majorana fermion DM.}
\label{P-MF}
\end{figure}

\begin{figure}[h]
\begin{center}
\minigraph{7cm}{-0.15in}{(a)}{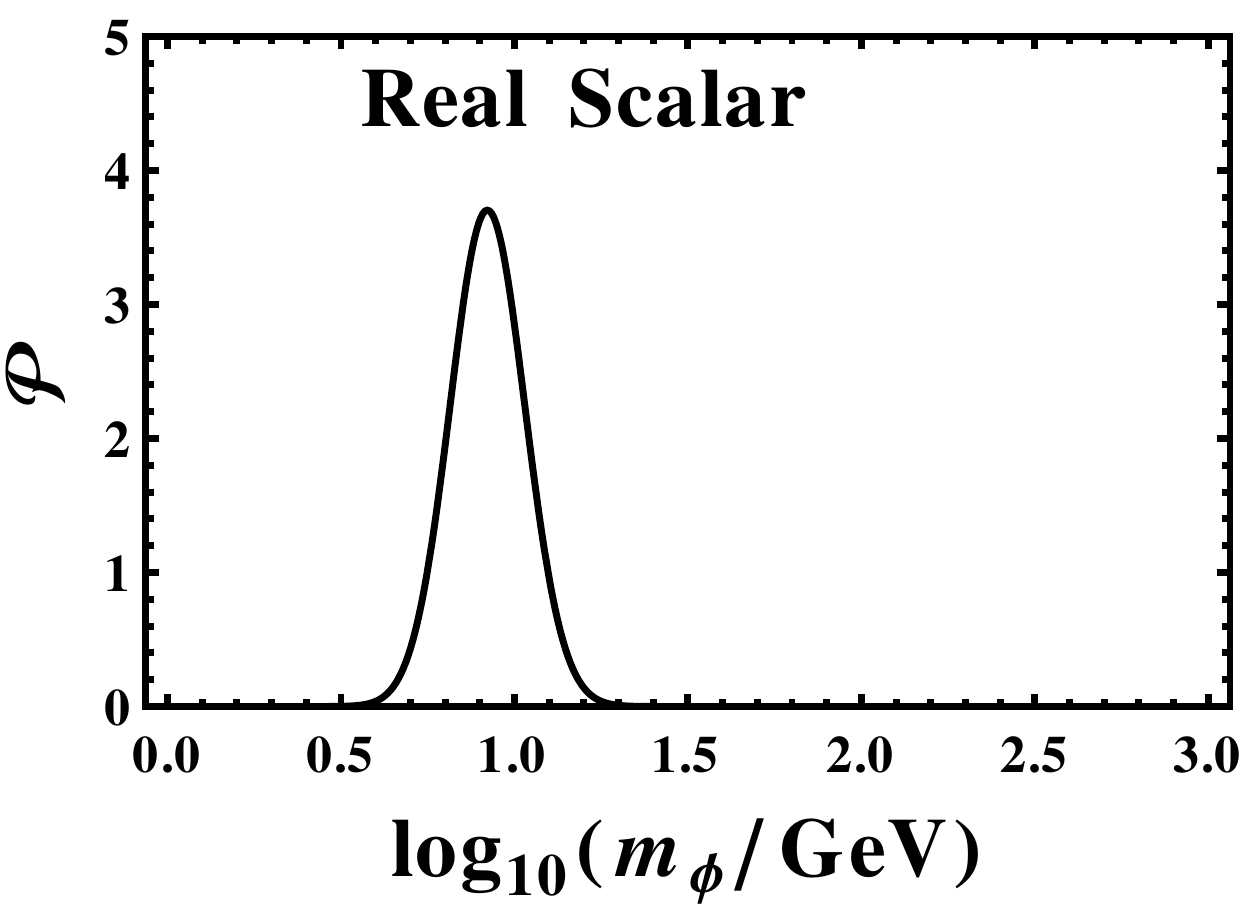}
\minigraph{7cm}{-0.15in}{(b)}{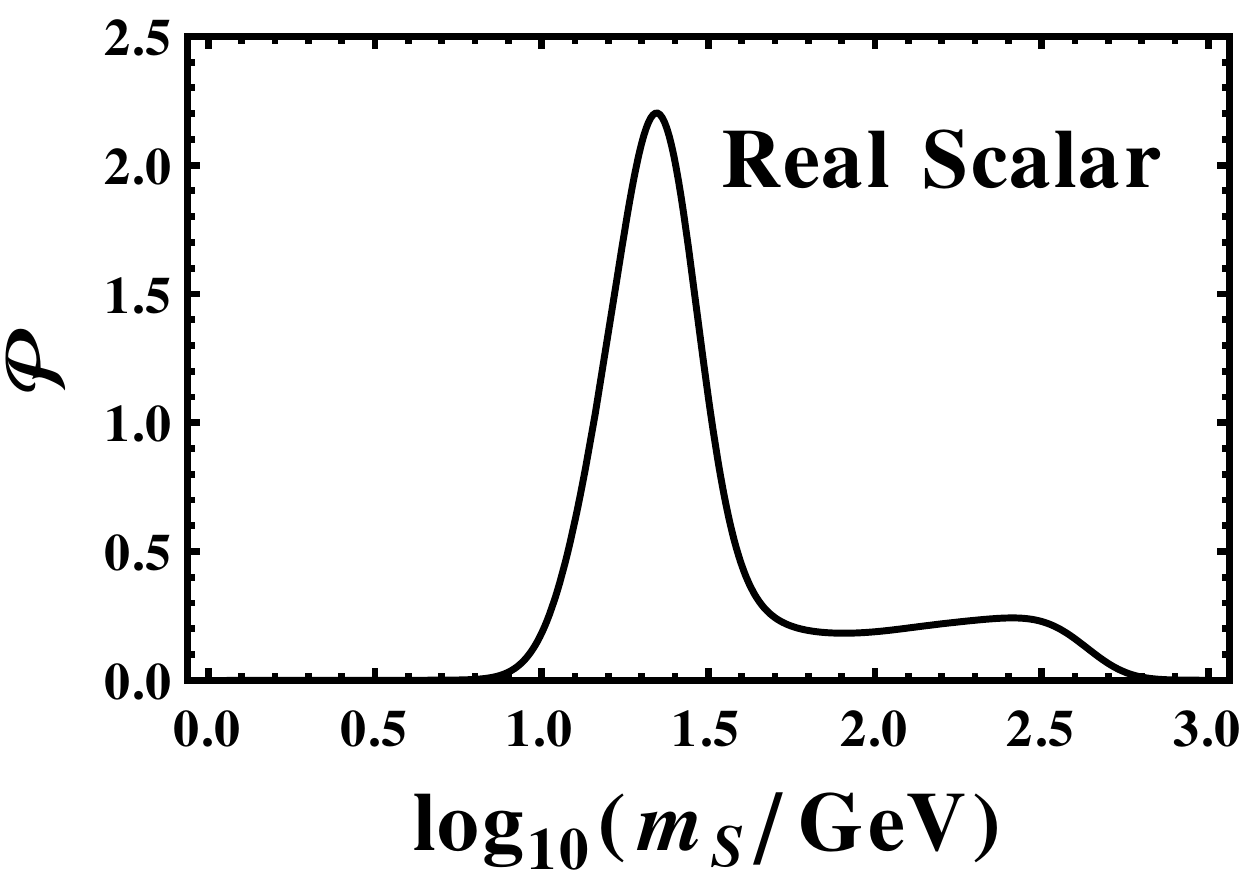}\\
\minigraph{7cm}{-0.15in}{(c)}{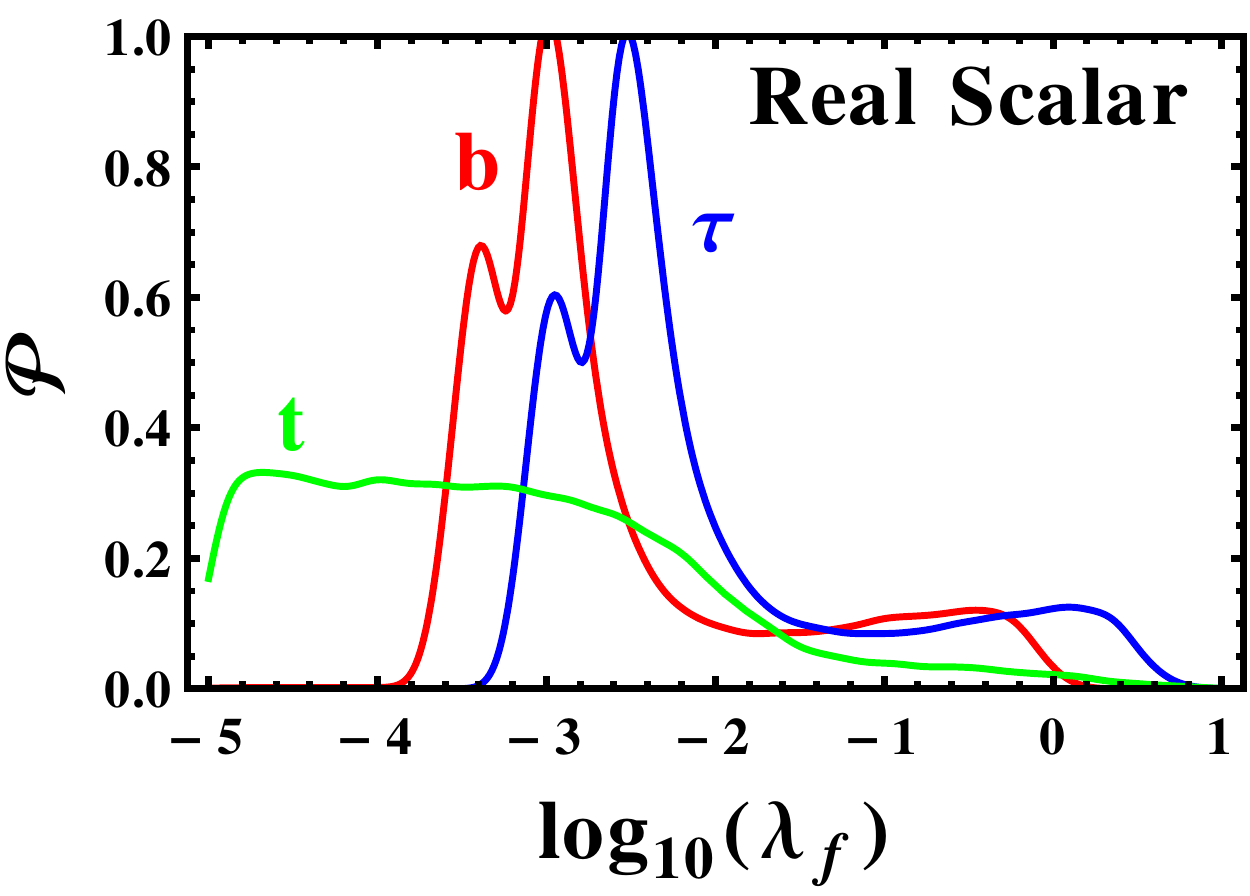}
%\minigraph{7cm}{-0.15in}{(c)}{plots/mchims-MF.pdf}
%\minigraph{7cm}{-0.15in}{(d)}{plots/SI-MF.pdf}
%\includegraphics[scale=1,width=7cm]{plots/btau-MF.pdf}
%\includegraphics[scale=1,width=7cm]{plots/bt-MF.pdf}\\
%\includegraphics[scale=1,width=7cm]{plots/mchims-MF.pdf}
%\includegraphics[scale=1,width=7cm]{plots/SI-MF.pdf}
\end{center}
\caption{Posterior probability distribution as a function of (a) $m_\phi$, (b) $m_S$ and (c) $\lambda_f$ ($f=b,t,\tau$) for real scalar DM.}
\label{P-RS}
\end{figure}

\begin{figure}[h]
\begin{center}
\minigraph{7cm}{-0.15in}{(a)}{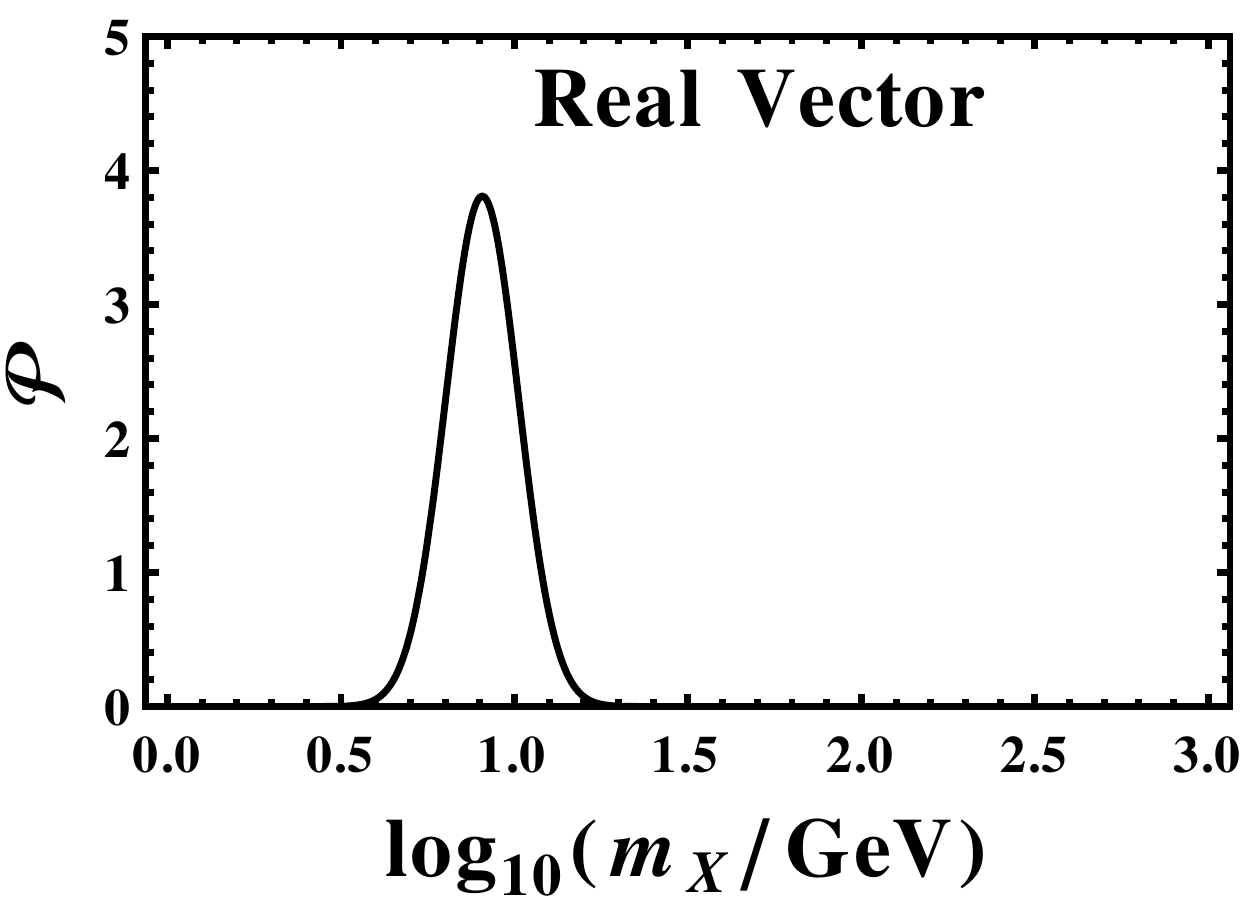}
\minigraph{7cm}{-0.15in}{(b)}{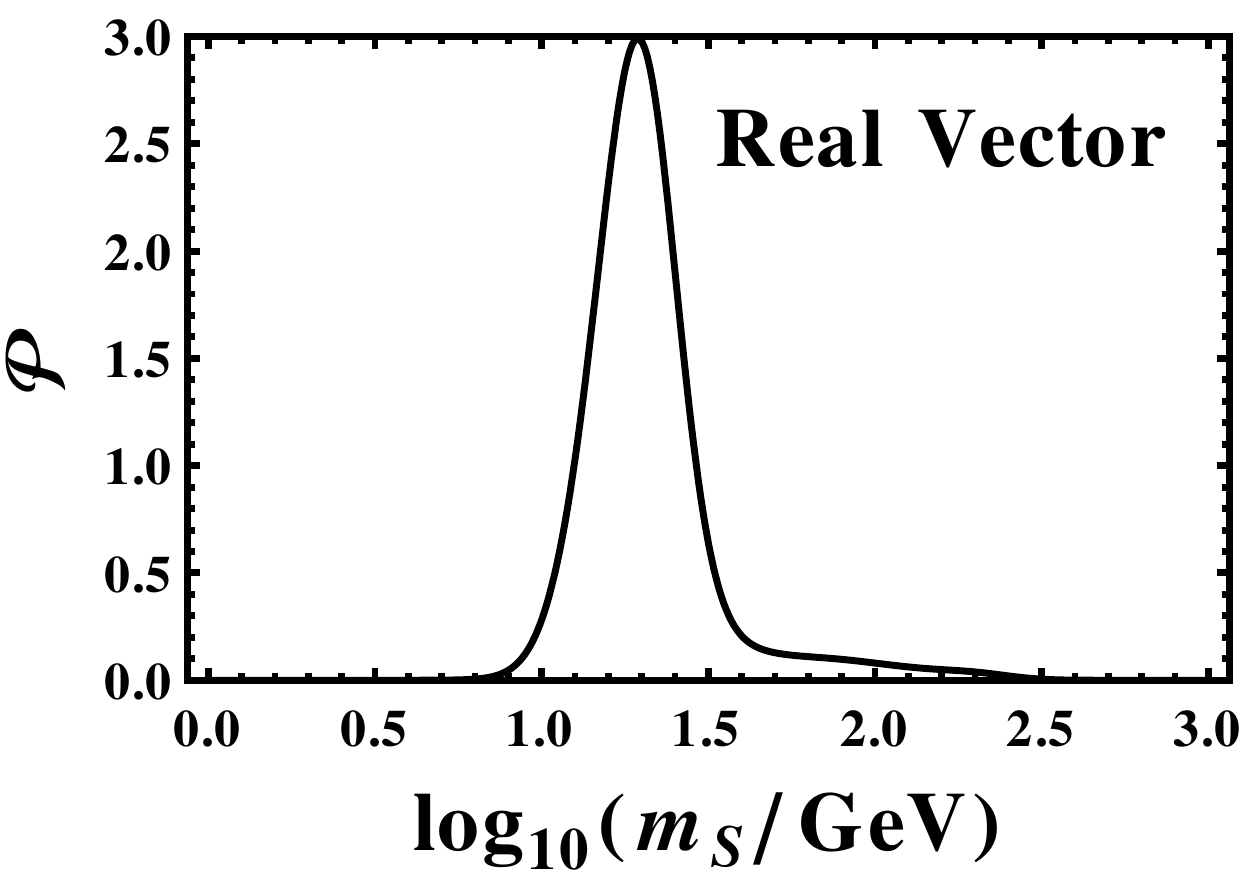}\\
\minigraph{7cm}{-0.15in}{(c)}{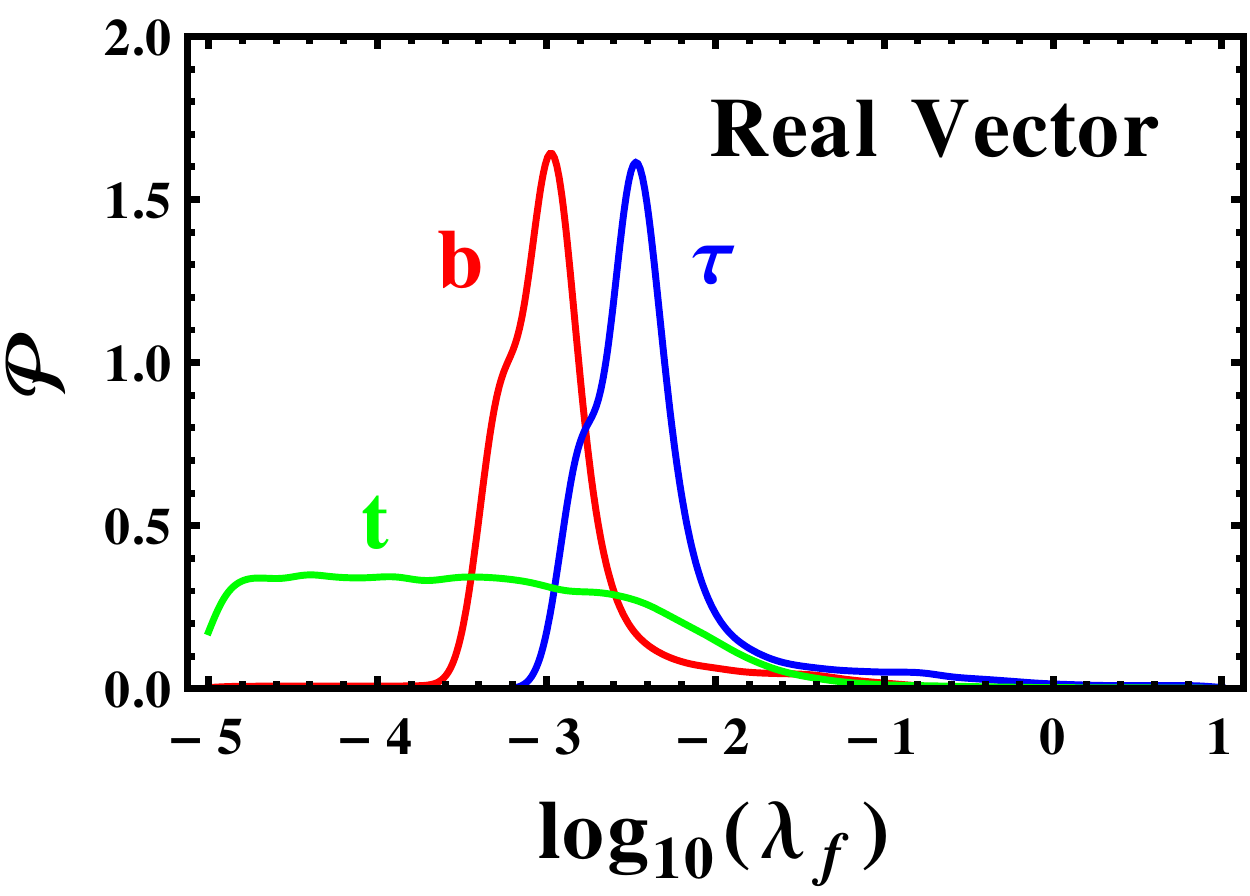}
%\minigraph{7cm}{-0.15in}{(c)}{plots/mchims-MF.pdf}
%\minigraph{7cm}{-0.15in}{(d)}{plots/SI-MF.pdf}
%\includegraphics[scale=1,width=7cm]{plots/btau-MF.pdf}
%\includegraphics[scale=1,width=7cm]{plots/bt-MF.pdf}\\
%\includegraphics[scale=1,width=7cm]{plots/mchims-MF.pdf}
%\includegraphics[scale=1,width=7cm]{plots/SI-MF.pdf}
\end{center}
\caption{Posterior probability distribution as a function of (a) $m_X$, (b) $m_S$ and (c) $\lambda_f$ ($f=b,t,\tau$) for real vector DM.}
\label{P-RV}
\end{figure}

In Figs.~\ref{P-MF}, \ref{P-RS} and \ref{P-RV}, we show the posterior probability distributions as a function of (a) DM mass, (b) $m_S$ and (c) $\lambda_f$ ($f=b,t,\tau$) for the Majorana fermion, real scalar and real vector dark matter models, respectively.  One can see that the preferred dark matter mass is in a narrow region around some certain value for all three candidates, that is $m_{\chi}\simeq 35$ GeV, $m_{\phi,X}\simeq 8$ GeV.  For scalar and vector dark matter a resonance funnel value of the mediator mass is favored, that is $m_S\simeq 2 m_{\phi,X}$, while its distribution has bimodal structure for the fermionic case, $m_S\lesssim 10$ GeV or $m_S\gtrsim 100$ GeV.  Amongst the mediator to standard fermions couplings $\lambda_b$ is strongly favored for Majorana fermion dark matter.  In the real scalar and vector cases the tau lepton final state has a comparable contribution to that of the b quark. %as smaller DM mass is favored in this case.
The bimodal structure of the mediator mass in the fermionic case is due to the relatively large coupling to b quark $\lambda_b\simeq 10^{-2}$ compared to $\lambda_b\simeq 10^{-3}$ in the other two models.  The smaller $\lambda_b$ values, and the favored resonance funnel value of $m_S$ for the scalar and vector models, are induced by the requirement of evading direct detection limits as the cross sections are not suppressed in these two cases.

\begin{figure}[h]
\begin{center}
\minigraph{7cm}{-0.15in}{(a)}{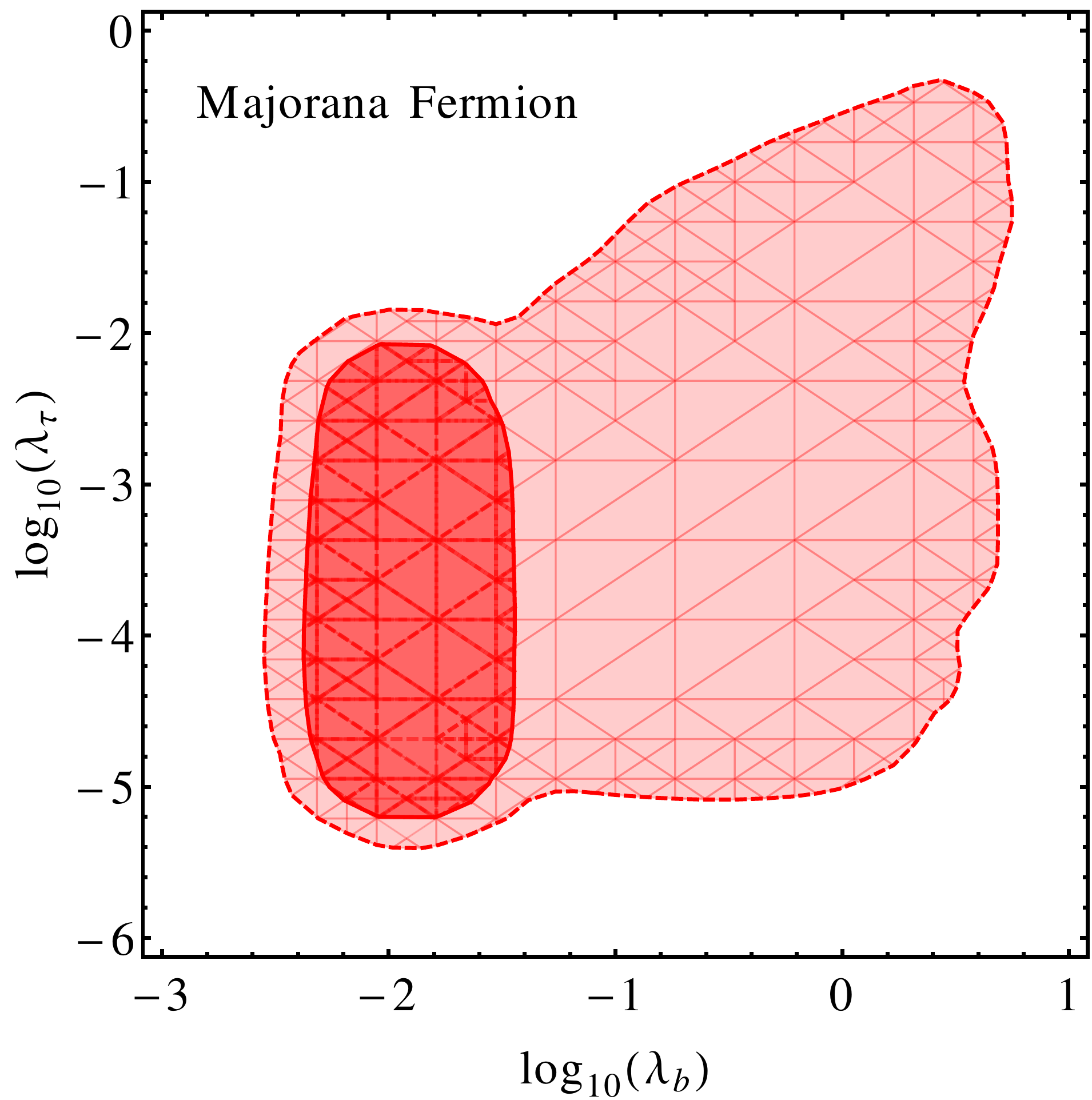}
\minigraph{7cm}{-0.15in}{(b)}{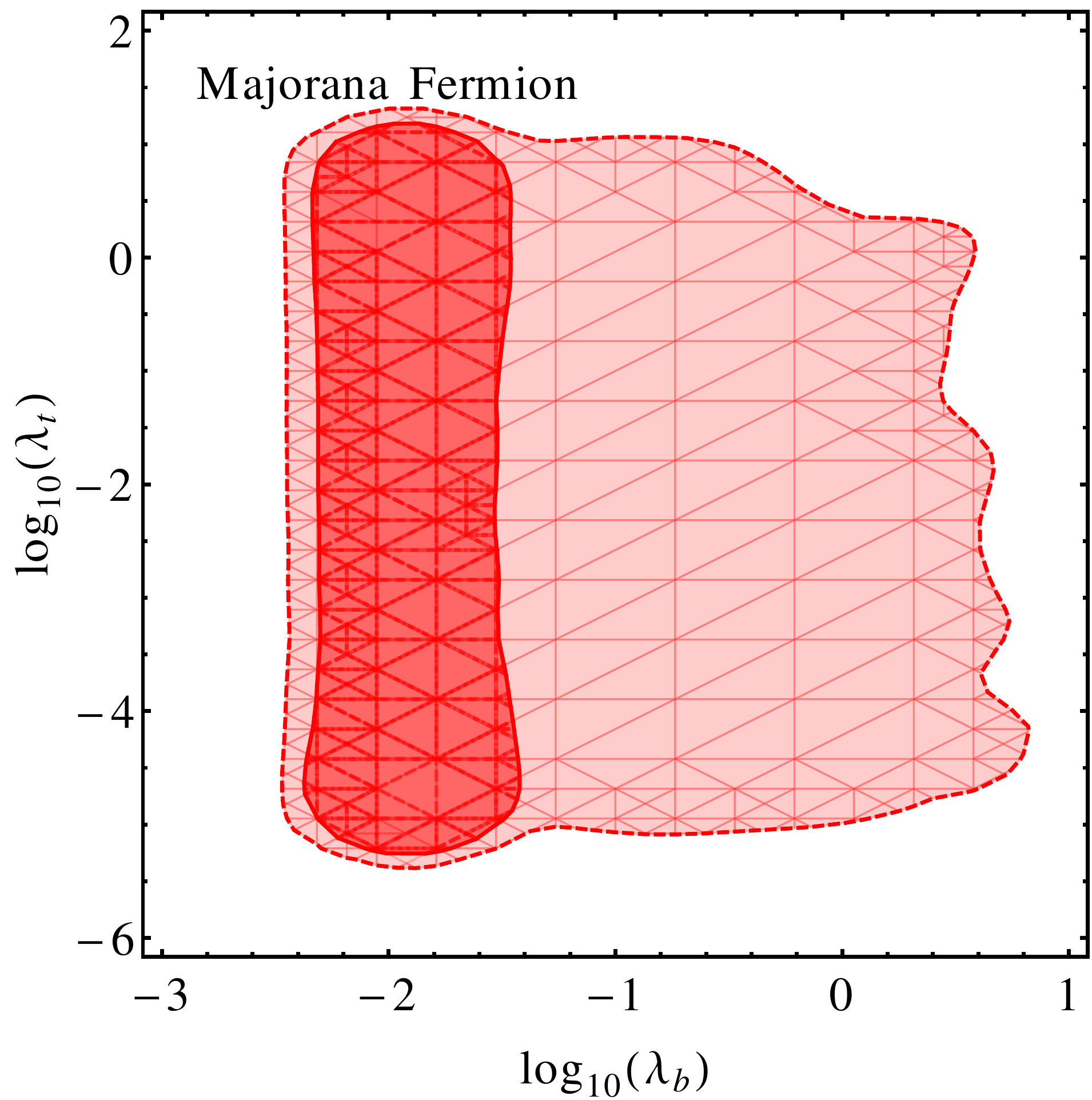}\\
\minigraph{7cm}{-0.15in}{(c)}{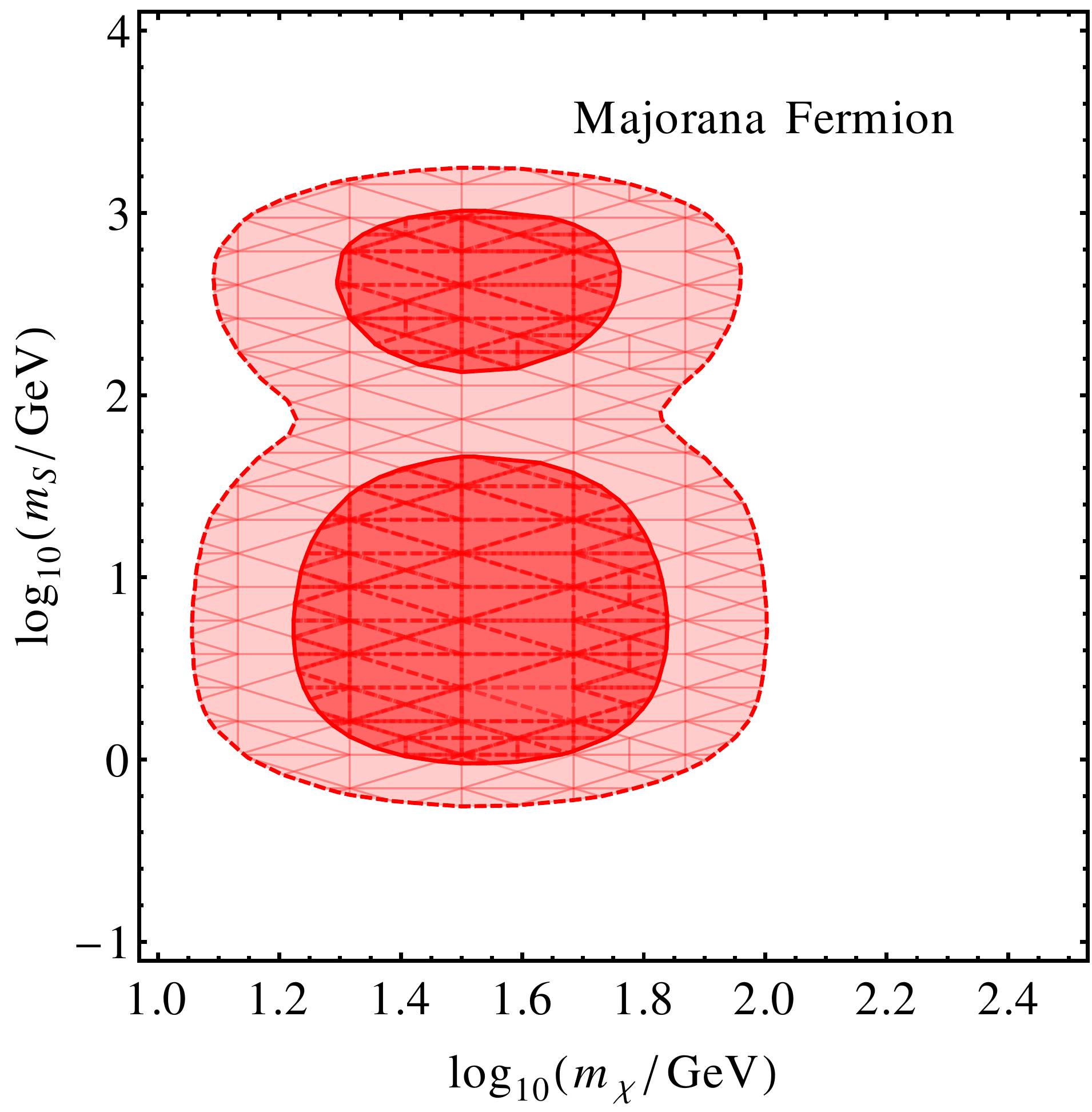}
\minigraph{7cm}{-0.15in}{(d)}{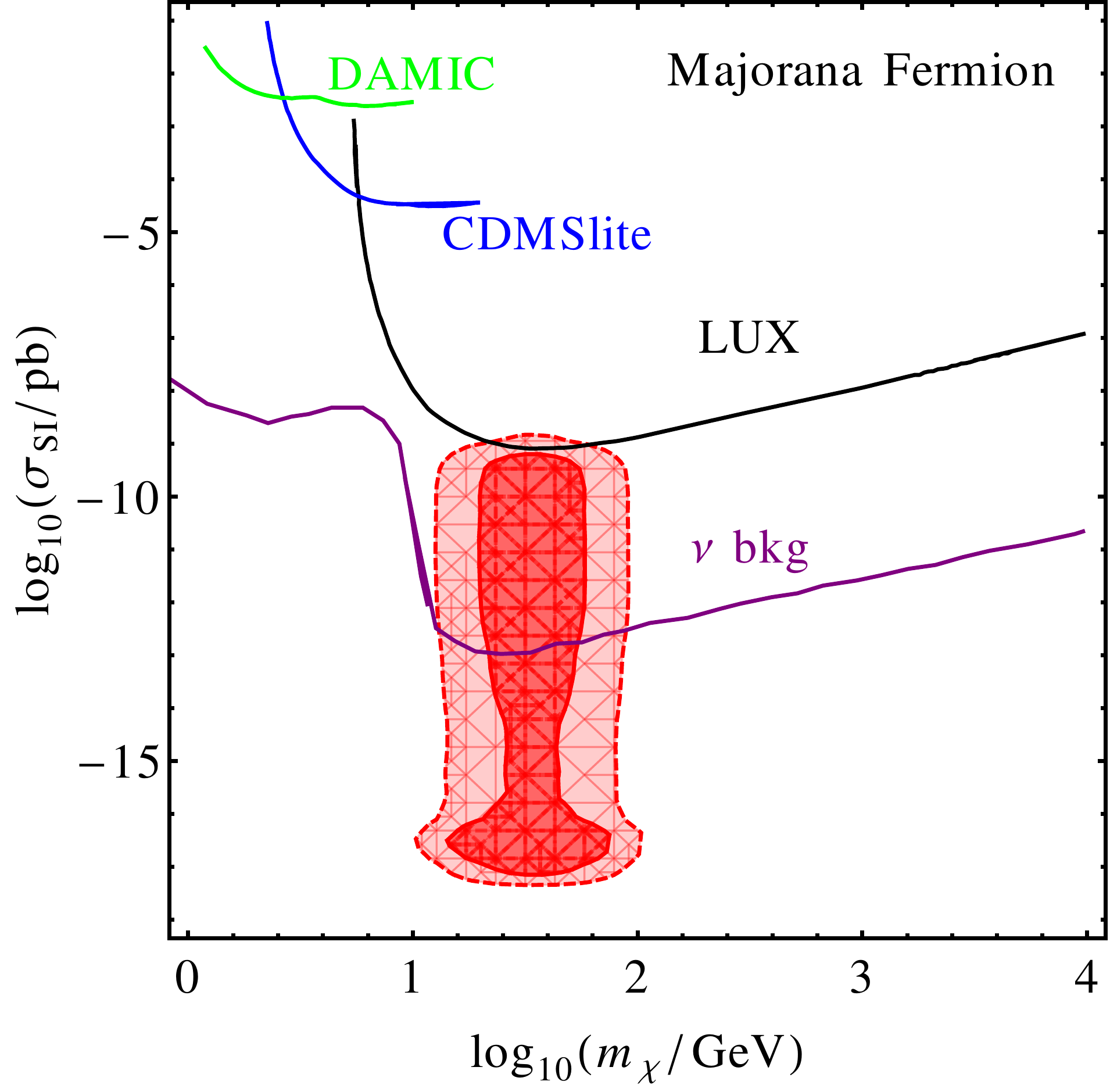}
\end{center}
\caption{Posterior probability distribution marginalized to (a) $\lambda_\tau$ vs. $\lambda_b$, (b) $\lambda_t$ vs. $\lambda_b$, (c) $m_S$ vs. $m_\chi$ and (d) $\sigma_{\rm SI}$ vs. $m_\chi$ for Majorana fermion DM. The light and dark regions correspond to 1 and 2$\sigma$ credible regions, respectively.}
\label{MF}
\end{figure}

\begin{figure}[h]
\begin{center}
\minigraph{7cm}{-0.15in}{(a)}{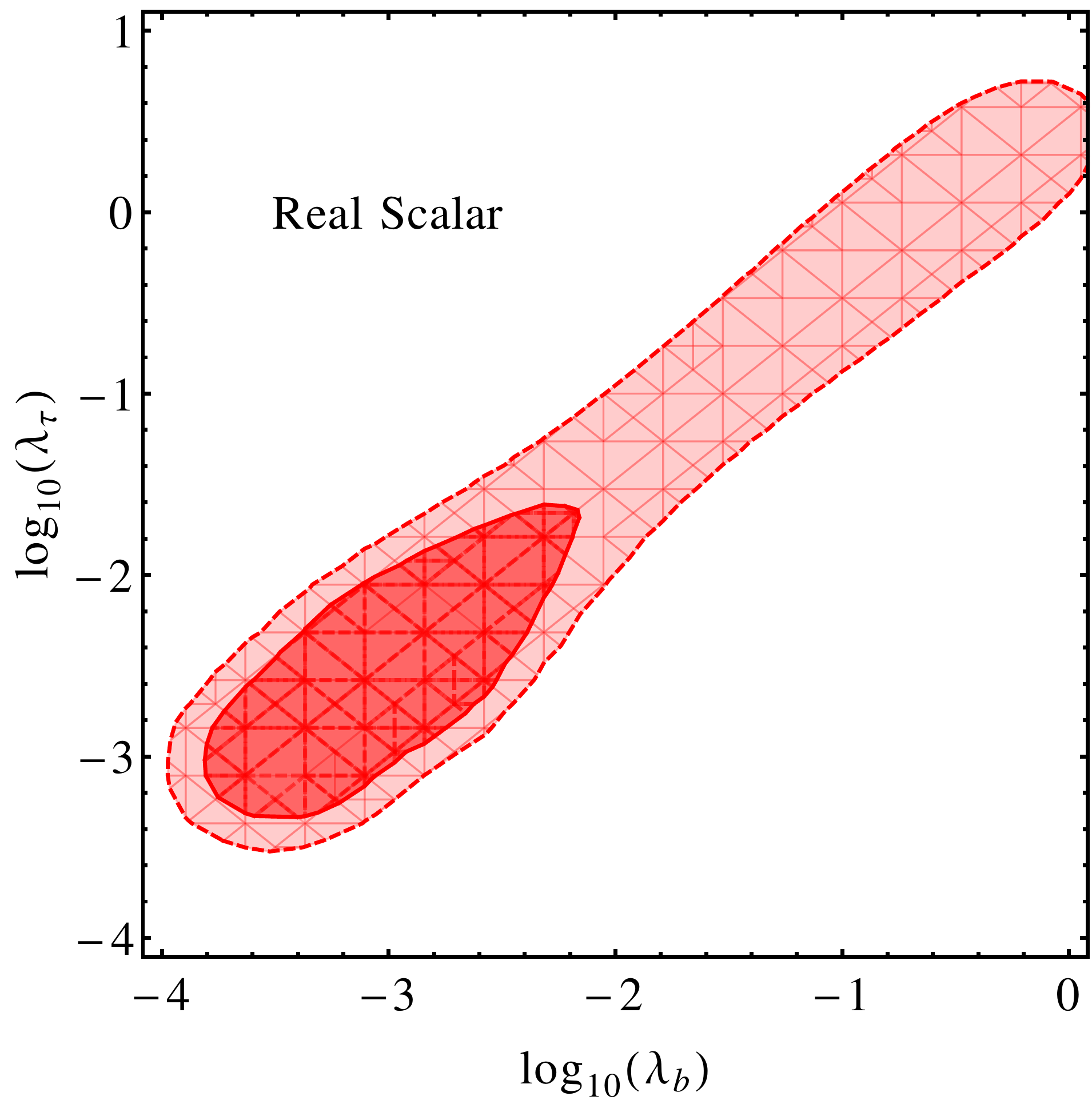}
\minigraph{7cm}{-0.15in}{(b)}{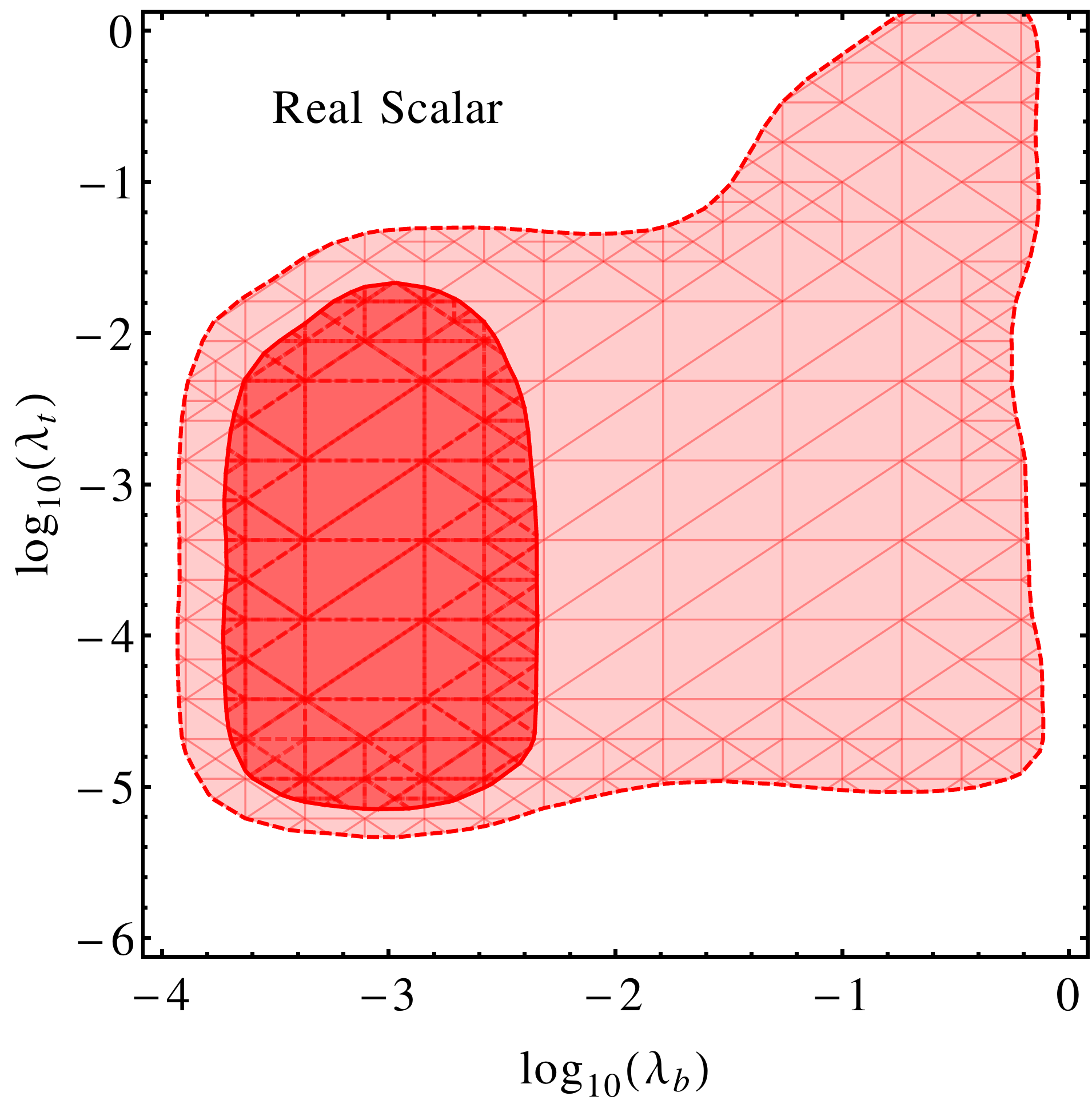}\\
\minigraph{7cm}{-0.15in}{(c)}{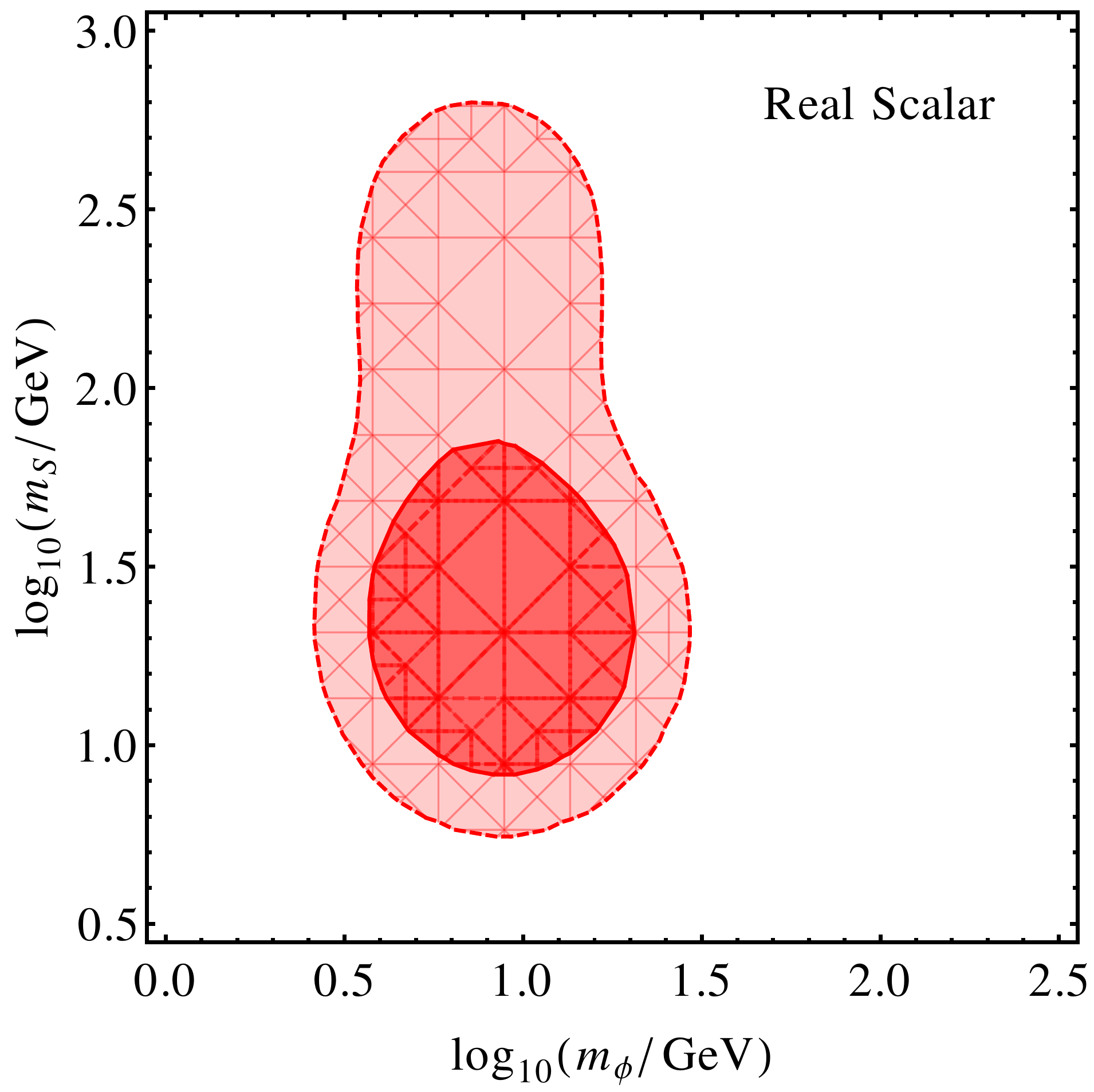}
\minigraph{7cm}{-0.15in}{(d)}{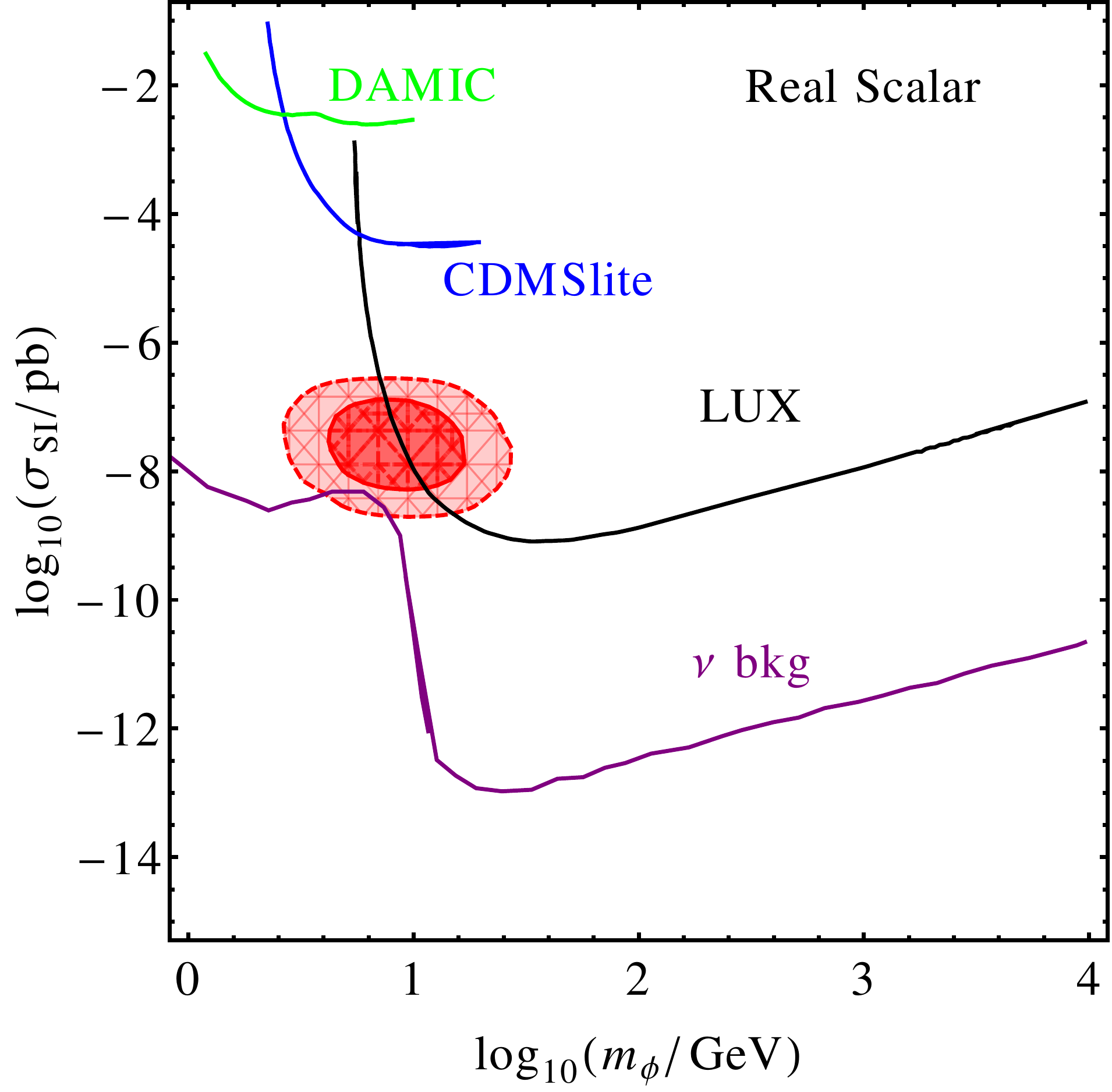}
\end{center}
\caption{Posterior probability distribution marginalized to (a) $\lambda_\tau$ vs. $\lambda_b$, (b) $\lambda_t$ vs. $\lambda_b$, (c) $m_S$ vs. $m_\phi$ and (d) $\sigma_{\rm SI}$ vs. $m_\phi$ for real scalar DM.}
\label{RS}
\end{figure}

\begin{figure}[h]
\begin{center}
\minigraph{7cm}{-0.15in}{(a)}{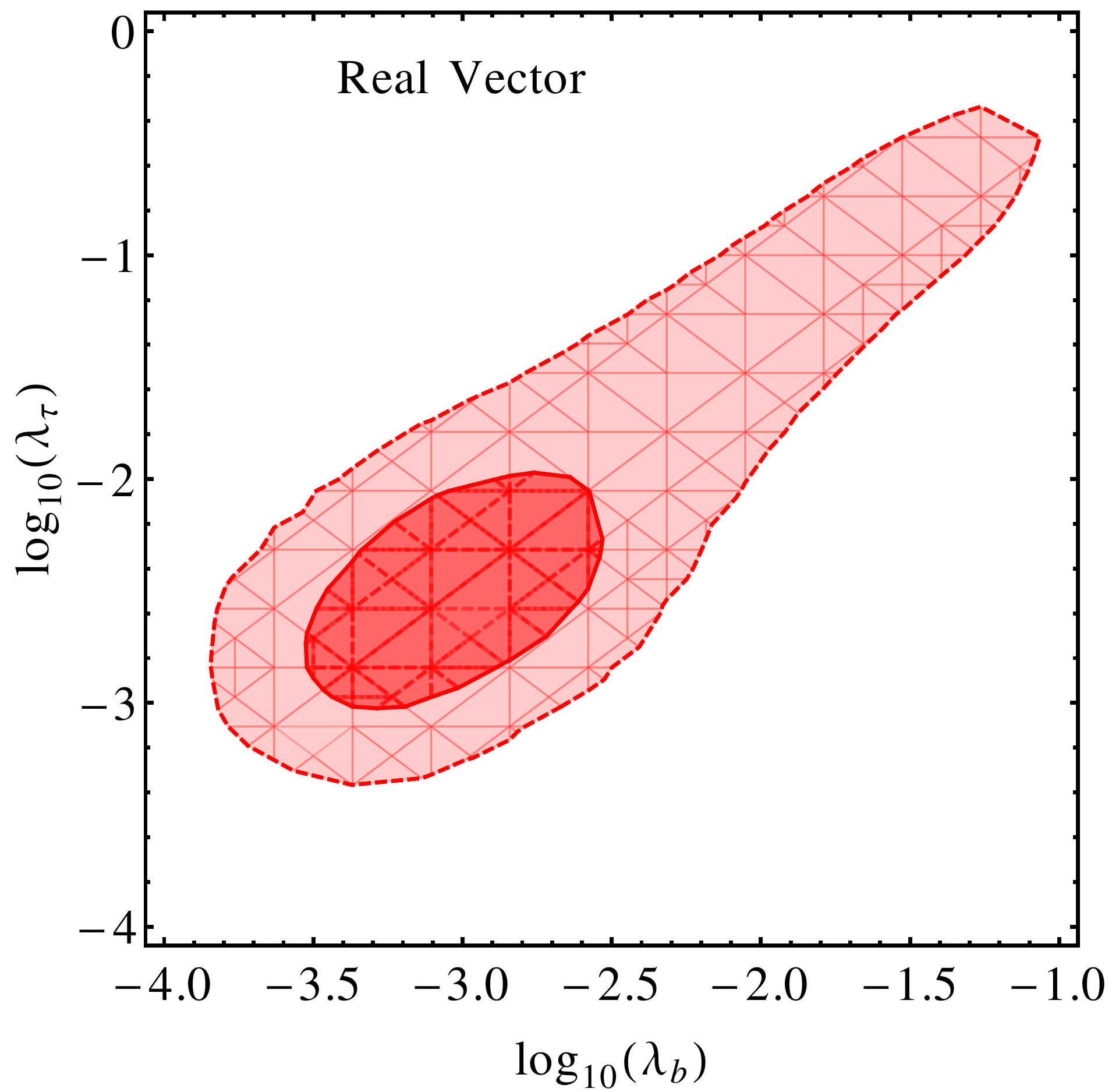}
\minigraph{7cm}{-0.15in}{(b)}{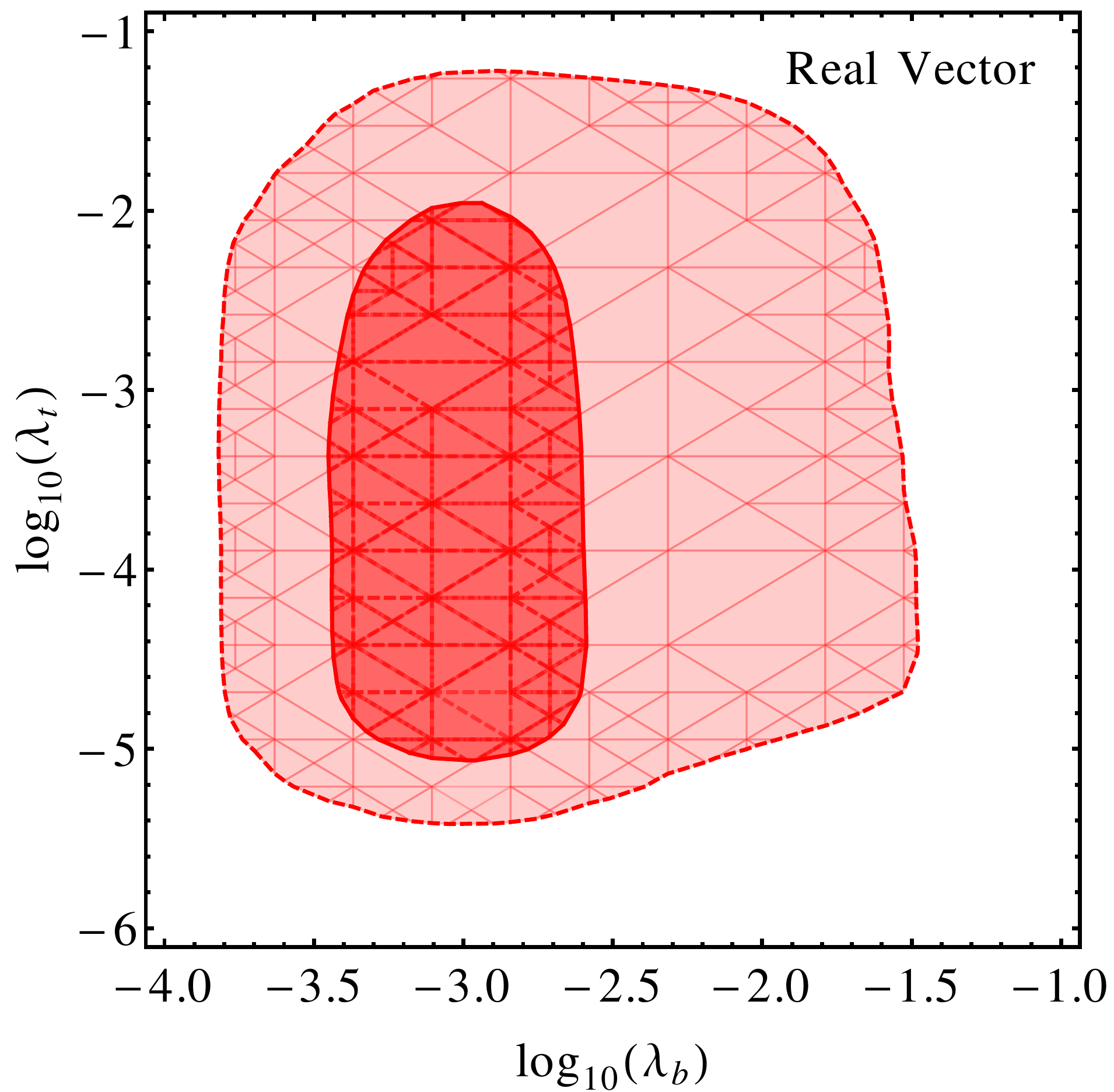}\\
\minigraph{7cm}{-0.15in}{(c)}{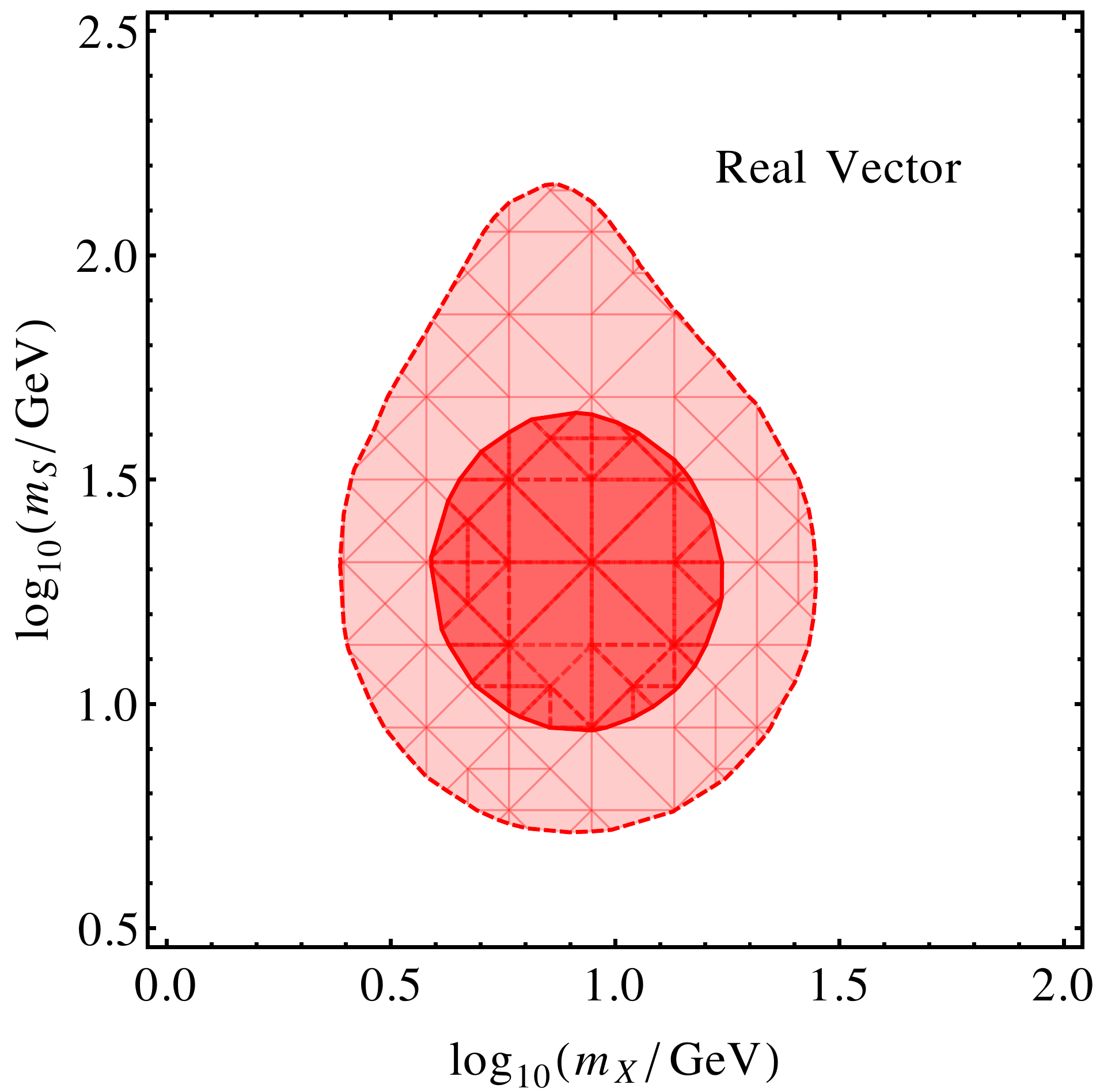}
\minigraph{7cm}{-0.15in}{(d)}{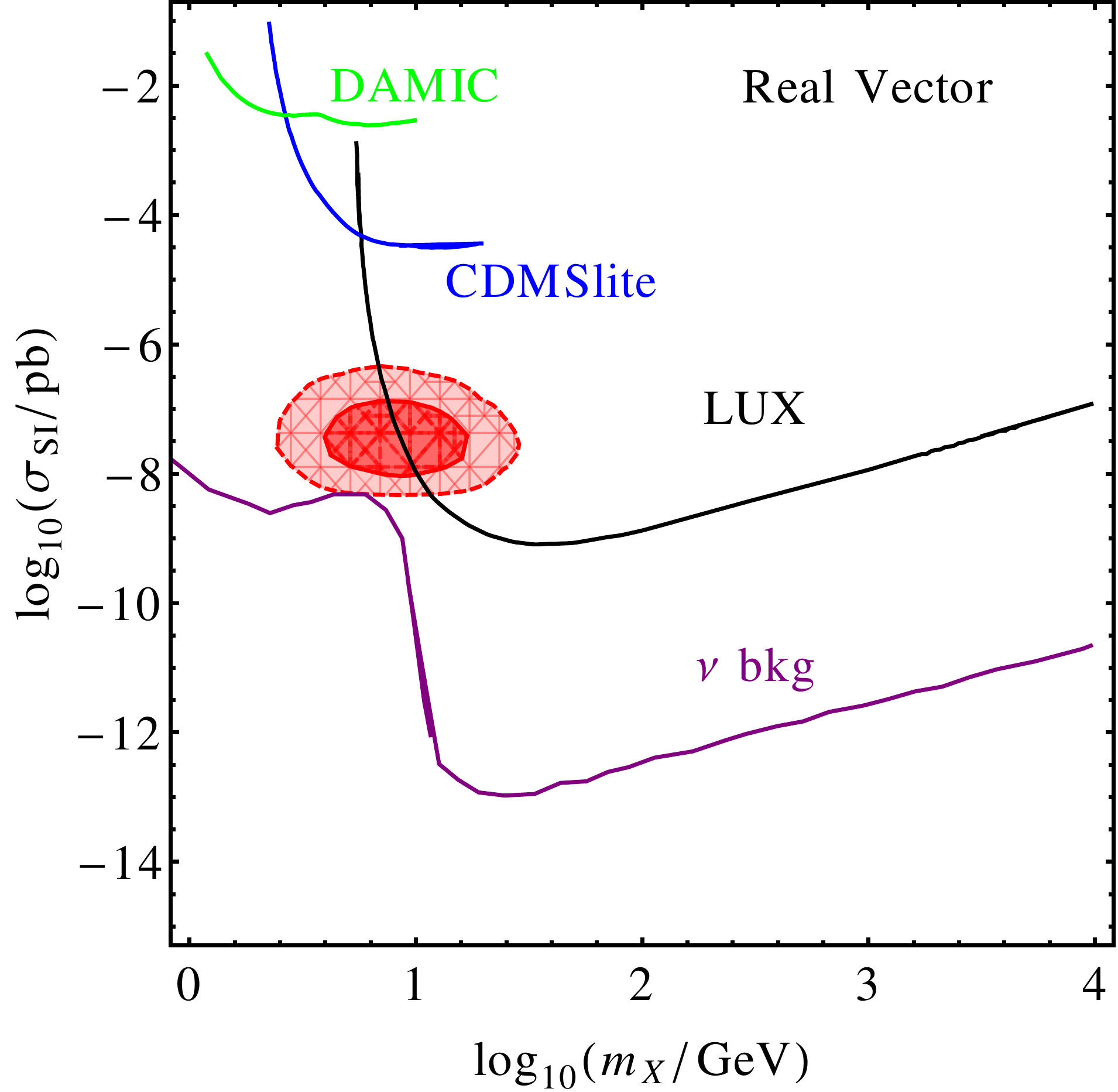}
\end{center}
\caption{Posterior probability distribution marginalized to (a) $\lambda_\tau$ vs. $\lambda_b$, (b) $\lambda_t$ vs. $\lambda_b$, (c) $m_S$ vs. $m_X$ and (d) $\sigma_{\rm SI}$ vs. $m_X$ for real vector DM.}
\label{RV}
\end{figure}

After marginalizing the posterior probability to two model parameters, we obtain the 1 and 2$\sigma$ credible regions shown in Figs.~\ref{MF}, \ref{RS} and \ref{RV} for the three models.  For Majorana fermion dark matter, we obtain $0.0025(0.004)\lesssim\lambda_b\lesssim 6(0.04)$, $\lambda_\tau\lesssim 1.6(0.01)$ at 1(2)$\sigma$ credible level.  As the dark matter mass $m_\chi$ is favored to be in the region $10(15)-100(70) \ {\rm GeV}<m_t$ at the 1(2)$\sigma$ credible level, and the elastic scattering cross section with nuclei is velocity-suppressed (c.f. Eq.~(\ref{DDMF})), the $t\bar{t}$ channel is forbidden and the direct detection limits do not affect this scenario.  Consequently $\lambda_t$ is not constrained in this case.  The mediator mass has two favored 2$\sigma$ regions: $10^2\lesssim m_S\lesssim 10^3$ GeV and $1\lesssim m_S\lesssim 40$ GeV, while the whole range of $1<m_S<10^3$ GeV is available at the 1$\sigma$ credible level.  The two favored regions of $m_S$ give two spin-independent scattering cross section regions evading the direct detection limits for fermionic dark matter, as seen in Fig.~\ref{MF} (d).  The heavy mediator masses, unfortunately, lead to $\sigma_{SI}$ partly lower than neutrino background.

For the real scalar dark matter model, we obtain $0.0001(0.00015)\lesssim\lambda_b\lesssim 1(0.01)$, $0.00025(0.0004)\lesssim\lambda_\tau\lesssim 6(0.025)$ at 1(2)$\sigma$ credible level.  As the scattering cross section on nuclei is sizable compared to the fermionic case and proportional to $\lambda_t^2$, the large $\lambda_t$ region is constrained by direct detection although the $t\bar{t}$ channel is not allowed, which results in $\lambda_t\lesssim 1(0.01)$ at 1(2)$\sigma$ credible level.  The dark matter and mediator mass, $m_\phi$ and $m_S$, both have single favored region: $2.5(4)\lesssim m_\phi\lesssim 32(20)$ GeV, $5(8)\lesssim m_S\lesssim 630(80)$ GeV at 1(2)$\sigma$ credible level. The favored region of spin-independent direct detection cross section is narrow, i.e. two orders of magnitude below the LUX limit.

In the real vector dark matter case, we have $0.0001(0.0003)\lesssim\lambda_b\lesssim 0.1(0.003)$, $0.0004(0.001)\lesssim\lambda_\tau\lesssim 0.6(0.01)$ at 1(2)$\sigma$ credible level.  For the same reason as in the scalar case, the top quark coupling stays in the favored region $\lambda_t\lesssim 0.1(0.01)$ at 1(2)$\sigma$ credible level.  The dark matter and mediator mass, $m_X$ and $m_S$, are favored in lower regions than for the scalar model: $2.5(4)\lesssim m_X\lesssim 32(20)$ GeV, $5(10)\lesssim m_S\lesssim 160(40)$ GeV at 1(2)$\sigma$ credible level.  The small favored vector dark matter mass gives $\sigma_{SI}\simeq 10^{-8}-10^{-6}$ pb.

%\begin{figure}[htb]
%\begin{center}
%\hspace{9.20mm}
%\includegraphics[width=8cm,trim=5mm -7mm -9mm 0mm,clip=true,height=20mm]{newplots/P-mchi-MF.pdf} \\
%\vspace{-3.9mm}
%\includegraphics[width=8cm]{newplots/mchims-MF.pdf}
%\hspace{-4.6mm}
%\includegraphics[width=7.11cm,trim=-6mm 10mm -8.mm 10mm,clip=true,angle=-90,origin=r,height=50mm]{newplots/P-ms-MF.pdf}
%\includegraphics[scale=1,width=7cm]{newplots/P-mchi-MF.pdf}
%\end{center}
%\caption{Majorana fermion. The light and dark regions correspond to 1 and 2$\sigma$ credible regions, respectively.}
%\label{fig1}
%\end{figure}

Using Bayesian statistics, we can compare the relative viability of our dark matter models.  We use Bayes factors, defined as the ratio of the model evidences, to perform such a model comparison.  In Table~\ref{bayes} we show the natural logarithm of the Bayes factors for the dark matter models we consider.  According to Jeffreys' scale~\cite{Arina:2013jma}, the Majorana fermion model is moderately favored over the vector model, which in turn is strongly favored against the scalar model.

\begin{table}[h]
\begin{tabular}{|c|c|c|c|}
\hline
${\rm ln}(\frac{{\rm evidence}_A}{{\rm evidence}_B})$ & $\chi_B$ & $\phi_B$ & $X_B$ \\ \hline
$\chi_A$ & $0$ & +10.0 & +2.60  \\ \hline
$\phi_A$ & $-$ &   $0$ & -7.40  \\ \hline
   $X_A$ & $-$ &   $-$ & $0$    \\ \hline
\end{tabular}
\caption{The natural logarithm of model evidence ratios (Bayes factors) for the three dark matter models we considered: real scalar $\phi$, Majorana fermion $\chi$, and real vector $X$.}
\label{bayes}
\end{table}

Finally, we discuss the LHC constraint on the Majorana fermion dark matter model. According to Fig.~\ref{MF} (c) the minimal condition of applying the LHC limit i.e. $m_S>2m_\chi$ gives $\sim 100$ GeV of minimal $m_S$ in the favored region which is smaller than the missing energy cut $\cancel{E}_T^{cut}$.  Together with the favored coupling $\lambda_b,\lambda_t$ we obtain the minimal $M_\ast^{\rm simp,f}$:
\begin{eqnarray}
M_\ast^{\rm simp,b}>\sqrt[3]{350^2\cdot m_b\over 1\cdot 10^{-2}}\simeq 380 \ {\rm GeV}, \ \ \ M_\ast^{\rm simp,t}>\sqrt[3]{225^2\cdot m_t\over 1\cdot 1}\simeq 205 \ {\rm GeV}.
\end{eqnarray}
They are both larger than the maximal lower limit on cut-off scale $M_\ast^{\rm limit,b}\simeq 100 (130)$ GeV and $M_\ast^{\rm limit,t}\simeq 110 (130)$ GeV obtained at the 8 (14) TeV LHC~\cite{Lin:2013sca}.  We also include the LHC limit in our numerical calculation and find our results indeed do not change.
Thus, we can conclude that LHC limit does not give stringent constraint on our model.
%Finally, we show the fermionic dark matter results taking into account of the collider requirement in Eq.~(\ref{lhc}). xxxxxxxxxxxx

%\begin{figure}[t]
%\begin{center}
%\includegraphics[scale=1,width=7cm]{plots/btau-MFwcollider.pdf}
%\includegraphics[scale=1,width=7cm]{plots/bt-MFwcollider.pdf}\\
%\includegraphics[scale=1,width=7cm]{plots/mchims-MFwcollider.pdf}
%\includegraphics[scale=1,width=7cm]{plots/SI-MFwcollider.pdf}
%\end{center}
%\caption{Majorana fermion with collider limit.}
%\label{fig1}
%\end{figure}

%%%%%%%%%%%%%%%%%%%%%%%%%%%%%%%%%%%%%%%%%%%%%%%%%%%%%%%%%%%%%%%%
\section{Conclusions}
\label{sec:Concl}
%%%%%%%%%%%%%%%%%%%%%%%%%%%%%%%%%%%%%%%%%%%%%%%%%%%%%%%%%%%%%%%%

We use the stretched effective operators to introduce spin-0 mediating particle in general dark matter models with Majorana fermion, real scalar and vector candidates. We then sum over the contributions of the third generation SM fermions into gamma ray spectrum in the calculation of differential gamma ray flux.  In our analysis we use the gamma-ray data points shown in the right frame of Fig. 5 in Ref.~\cite{Daylan:2014rsa}.  As a result the gamma ray data strictly constrains our model parameters and dark matter mass.
We also calculate dark matter relic density, elastic scattering cross section and effective collider limit and contrast them with experiments. Using Bayesian inference we confine the most fundamental properties of dark matter particles, such as their mass and interaction
strength with ordinary matter and force particles.

We find
\begin{itemize}
\item the combination of gamma ray data, relic density and direct detection set stringent constraint on the three models we consider but the search for dark matter particle at the LHC does not;
\item posterior probability distributions show that the favored DM mass is in a narrow region around some certain value: $m_{\chi}\simeq 35$ GeV, $m_{\phi,X}\simeq 8$ GeV. The mediator mass $m_S$ favors $m_S\simeq 2 m_{\phi,X}$ for scalar and vector DM, while for fermionic DM case: $m_S\lesssim 10$ GeV or $m_S\gtrsim 100$ GeV. The mediator-SM fermions coupling $\lambda_b$ is strongly favored for Majorana fermion ($\lambda_b\simeq 10^{-2}$) DM. In real scalar and vector cases, tau lepton has comparable contribution as b quark ($\lambda_b\simeq 10^{-3}$);
%    The bimodal structure of mediator mass in fermionic DM case is due to the relatively large coupling to b quark $\lambda_b\simeq 10^{-2}$ compared to $\lambda_b\simeq 10^{-3}$ in other two models.
\item favored regions of parameters and direct detection cross section in Majorana fermion DM model at 1(2)$\sigma$ credible level:

$0.0025(0.004)\lesssim\lambda_b\lesssim 6(0.04)$, $\lambda_\tau\lesssim 1.6(0.01)$, $\lambda_t$ unconstrained,

%$10(15)\lesssim m_\chi\lesssim 10^2(70)$ GeV, $1<m_S<10^3$ GeV ($10^2\lesssim m_S\lesssim 10^3$ GeV, $1\lesssim m_S\lesssim 25$ GeV),
$10(15)\lesssim m_\chi\lesssim 10^2(70)$ GeV, $m_S$ unconstrained ($10^2\lesssim m_S\lesssim 10^3$ GeV, $1\lesssim m_S\lesssim 40$ GeV),

$\sigma_{SI}\sim 10^{-18}-10^{-9}$ pb;
\item favored regions of parameters and direct detection cross section in real scalar DM model at 1(2)$\sigma$ credible level:

$0.0001(0.00015)\lesssim\lambda_b\lesssim 1(0.01)$, $0.00025(0.0004)\lesssim\lambda_\tau\lesssim 6(0.025)$, $\lambda_t\lesssim 1(0.01)$,

$2.5(4)\lesssim m_\phi\lesssim 32(20)$ GeV, $5(8)\lesssim m_S\lesssim 630(80)$ GeV,

$\sigma_{SI}\sim 10^{-8}-10^{-6}$ pb;
\item favored regions of parameters and direct detection cross section in real vector DM model at 1(2)$\sigma$ credible level:

$0.0001(0.0003)\lesssim\lambda_b\lesssim 0.1(0.003)$, $0.0004(0.001)\lesssim\lambda_\tau\lesssim 0.6(0.01)$, $\lambda_t\lesssim 0.1(0.01)$,

$2.5(4)\lesssim m_X\lesssim 32(20)$ GeV, $5(10)\lesssim m_S\lesssim 160(40)$ GeV,

$\sigma_{SI}\sim 10^{-8}-10^{-6}$ pb;
\item the Majorana fermion DM model has moderate evidence against the real vector DM one while the latter case is strongly favored against scalar DM model.
\end{itemize}

\acknowledgments
T.~L. would like to thank Benjamin Farmer, Samuel D.~McDermott, Jayden L.~Newstead and Chris Savage for helpful discussions.
C.~B. and T.~L. would also like to thank Tracy R. Slatyer for providing gamma ray data points in Ref.~\cite{Daylan:2014rsa}.
This work in part was supported by the ARC Centre of Excellence for Particle Physics at the Terascale.
The work of C.~B. and T.~L. is supported by the Australian Research Council.
%%%%%%%%%%%%%%%%%%%%%%%%%%%%%%%%%%%%%%%%%%%%
\appendix

%%%%%%%%%%%%%%%%%%%%%%%%%%%%%%%%%%%%%%%%%%%%
\section{Bayesian Inference}
%%%%%%%%%%%%%%%%%%%%%%%%%%%%%%%%%%%%%%%%%%%%

In this section we summarize the statistical background of our analysis.  Let $P(A|I)$ and $P(B|I)$ denote the plausibility of two non-exclusive propositions, $A$ and $B$, in light of some prior information, $I$.  The probability that both $A$ and $B$ are correct is given by the conditional expression \begin{equation}
 P(AB|I) = P(A|BI)P(B|I) .
\end{equation}
Bayes theorem follows from the symmetry of the conditional probability under the exchange of $A$ and $B$:
\begin{equation}
 P(A|BI) = \frac{P(B|AI)P(A|I)}{P(B|I)} .
\end{equation}
In this context $P(A|I)$ is typically called the prior probability and represents the plausibility of our hypothesis given the information prior to the observation of $B$.  The likelihood function $P(B|AI)$ indicates how accurately the hypothesis can replicate the data.  The posterior probability $P(A|BI)$ quantifies the plausibility of the hypothesis $A$ given the data $B$.  The evidence $P(B|I)$ serves to normalize the posterior.

For theoretical models with a continuous parameter $\theta$ Bayes' theorem can be recast in the form
\begin{equation}
\mathcal{P}(\theta|B,I) = \frac{\mathcal{L}(B|\theta,I)\pi(\theta,I)}{\epsilon(B,I)} .
\label{eqnBayes}
\end{equation}
The posterior distribution can be used to estimate the most likely region of $\theta$.  The evidence is calculated via an integral over the full parameter space
\begin{equation}
\epsilon(B,I) = \int_\theta \mathcal{L}(B|\theta,I)\pi(\theta,I) d\theta .
\end{equation}
For more than one continuous parameters, $\theta_i$, marginalization is performed by integrating the posterior over various parameters in the higher dimensional parameter space
\begin{equation}
\mathcal{P}(\theta_j) = \int \prod_{i\neq j} d\theta_i \mathcal{P}(\theta_i).
\end{equation}

%%%%%%%%%%%%%%%%%%%%%%%%%%%%%%%%%%%%%%%%
\section{Likelihood Functions}
%%%%%%%%%%%%%%%%%%%%%%%%%%%%%%%%%%%%%%%%%

Whenever an experimental central value is available with an uncertainty, we cast the likelihood function in the form of a Gaussian distribution centered on the measured value with standard deviation equal to the uncertainty:
\begin{equation}
\mathcal{L}_i(d|\theta,I) =  \frac{1}{\sqrt{2\pi}\sigma}\mathrm{Exp}\left(-\frac{(x(\theta) - d)^2}{2\sigma^2}\right).
\end{equation}
For experiments that only place a bound on a particular parameter, the likelihood function takes the form of a complementary error function:
\begin{equation}
\mathcal{L}_i(d|\theta,I) =  \frac{1}{2}\mathrm{Erfc} \left(\pm \frac{x(\theta) - d}{\sigma}\right) ,
\end{equation}
where the $+$ ($-$) sign is used for an upper (lower) limit.
The composite likelihood combines likelihood functions for various data points $d_i$ at the parameter point $\theta$:
\begin{equation}
\mathcal{L}(D|\theta,I) = \prod_i \mathcal{L}_i(d_i|\theta,I).
\end{equation}

We assume that the theoretical calculations of relic density and direct detection have a uniform uncertainty of 50\% and 10\% respectively throughout the whole parameter space. We stipulate a 10\% uncertainty of the collider limit and that the experimental and theoretical uncertainties are the same for gamma ray data and direct detection.  We combine experimental and theoretical uncertainties in quadrature.

%%%%%%%%%%%%%%%%%%%%%%%%%%%%%%%%%%%%%%%%%%%%%%%%%%%%%%%%%%%%%%%%%%%%
%\input{reference.tex}

\end{document}